\documentclass[aps,preprintnumbers,showpacs,nofootinbib]{revtex4}
\usepackage[english]{babel}
\usepackage{graphicx}
\usepackage{subfig}
\usepackage{hyperref}
\usepackage{dcolumn}
\usepackage{bm}
\usepackage{amsmath}
\newcommand{\newc}{\newcommand}
\newc{\ra}{\rightarrow}
\newc{\lra}{\leftrightarrow}
\newc{\beq}{\begin{equation}}
\newc{\eeq}{\end{equation}}
\newc{\barr}{\begin{eqnarray}}
\newc{\earr}{\end{eqnarray}}

\def\vbf{\mbox{\boldmath $\upsilon$}}
\def\barr{\begin{eqnarray}}
\def\earr{\end{eqnarray}}

 \def\vbf{\mbox{\boldmath $\upsilon$}}
  \def\Sbf{\mbox{\boldmath $\Sigma$}}

\def\bfs{\mbox{\boldmath $\sigma$}}
\begin{document}
\newcommand{\Od}{{\cal O}}
\newcommand{\lsim}   {\mathrel{\mathop{\kern 0pt \rlap
  {\raise.2ex\hbox{$<$}}}
  \lower.9ex\hbox{\kern-.190em $\sim$}}}
\newcommand{\gsim}   {\mathrel{\mathop{\kern 0pt \rlap
  {\raise.2ex\hbox{$>$}}}
  \lower.9ex\hbox{\kern-.190em $\sim$}}}

\title{ Theoretical direct WIMP detection rates for transitions to nuclear excited states.}

\author{J.D. Vergados$^{1,2}$, F.T.  Avignone III$^{2}$, P. Pirinen$^3$, P. C. Srivastava$^{4}$, 
M. Kortelainen$^{3}$ and  J. Suhonen$^3$ }
\affiliation{$^1$ Theoretical Physics,University of Ioannina, Ioannina, Gr 451 10, Greece}
\affiliation{$^{(2)}${\it University of South Carolina, Columbia, SC 29208, USA, }}
\affiliation{$^3$University of Jyvaskyla, Department of Physics, P.O. Box 35, FI-40014,
University of Jyvaskyla , Finland}
\affiliation{$^4$Department of Physics, Indian Institute of Technology, Roorkee 247667, India}
\vspace{0.5cm}
\begin{abstract}
The recent WMAP and Planck data have confirmed that exotic dark matter
together with the vacuum energy (cosmological constant) dominate
in the flat Universe. Many extensions of the standard model provide dark matter candidates, in particular Weakly Interacting Massive Particles (WIMPs).
 Thus the direct dark matter detection is central to
particle physics and cosmology.  Most of the research on this
issue has hitherto focused on the detection of the recoiling
nucleus. In this paper we study transitions to the excited states, possible in some nuclei, which have sufficiently low lying excited states. Examples considered previously were  the first excited states of $^{127}$I and $^{129}$Xe. We examine here $^{83}$Kr, which offers some kinematical advantages and is currently considered as a possible target.
We find appreciable branching ratios  for the inelastic scattering mediated by the spin cross sections, with an  inelastic event rate of $4.4\times 10^{-4}$kg$^{-1}$d$^{-1}$.
So, the extra signature of the gamma ray following the de-excitation of these states can, in principle,
be exploited experimentally. A brief discussion of the experimental feasibility is given.
\end{abstract}
\date{\today}
\maketitle
\section{Introduction}
The combined MAXIMA-1 \cite{MAXIMA-1}, BOOMERANG \cite{BOOMERANG},
DASI \cite{DASI} and COBE/DMR Cosmic Microwave Background (CMB)
observations \cite{COBE} imply that the Universe is flat
\cite{flat01}
and that most of the matter in
the Universe is Dark \cite{SPERGEL},  i.e. exotic. These results have been confirmed and improved
by the recent WMAP  \cite{WMAP06} and Planck \cite{PlanckCP13} data. Combining 
the data of these quite precise measurements one finds:
$$\Omega_b=0.0456 \pm 0.0015, \quad \Omega _{\mbox{{\tiny CDM}}}=0.228 \pm 0.013 , \quad \Omega_{\Lambda}= 0.726 \pm 0.015~$$
(the more  recent Planck data yield a slightly different combination $ \Omega _{\mbox{{\tiny CDM}}}=0.274 \pm 0.020 , \quad \Omega_{\Lambda}= 0.686 \pm 0.020)$. It is worth mentioning that both the WMAP and the Plank observations yield essentially the same value of $\Omega_m h^2$,
  but they differ in the value of $h$, namely $h=0.704\pm0.013$ (WMAP) and $h=0.673\pm0.012$ (Planck).
Since any ``invisible" non exotic component cannot possibly exceed $40\%$ of the above $ \Omega _{\mbox{{\tiny CDM}}}$
~\cite {Benne}, exotic (non baryonic) matter is required and there is room for cold dark matter candidates or WIMPs (Weakly Interacting Massive Particles).
Even though there exists firm indirect evidence for a halo of dark matter
in galaxies from the
observed rotational curves, see e.g. the review \cite{UK01}, it is essential to directly
detect such matter in order to 
unravel the nature of the constituents of dark matter. At present there exist many such candidates: the
LSP (Lightest Supersymmetric Particle) \cite{ref2a,ref2b,ref2c,ref2,ELLROSZ,Gomez,ELLFOR}, technibaryon \cite{Nussinov92,GKS06}, mirror matter\cite{FLV72,Foot11}, Kaluza-Klein models with universal extra dimensions\cite{ST02a,OikVerMou} etc. Additional theoretical tools are the structure of the nucleus, see e.g.  \cite{JDV06a,Dree00,Dree,Chen}, and the nuclear matrix elements \cite{Ress,DIVA00,JDV03,JDV04,VF07}.

In most
calculations the WIMP is supposed to be the  neutralino or LSP (lightest supersymmetric particle), which is assumed to be primarily a gaugino,
usually a bino. Models which predict a substantial fraction of
higgsino lead to a relatively large spin induced cross section due
to the Z-exchange. Such models tend to violate the LSP relic
abundance constraint and are not favored.  Some claims have
recently been made, however, to the effect that the WMAP relic
abundance constraint can be satisfied in the hyperbolic branch of
the allowed SUSY parameter space, even though the neutralino is
then primarily a higgsino \cite{CCN03}. We will not restrict ourselves to supersymmetry and we will  adopt the optimistic view that
the detection rates due to the spin may be large enough to be
exploited by the experiments, see, e.g., \cite{CHATTO,JDV03, WELLS} . Such a view is further
encouraged by the fact that, unlike the scalar interaction, the
axial current allows one to populate excited nuclear states,
provided that their energies are sufficiently low so that they are
accessible by the low energy LSP, a prospect proposed long time
ago \cite{GOODWIT} and considered in some detail by Ejiri and collaborators \cite{EFO93}. 

For a Maxwell-Boltzmann (M-B) velocity distribution 
the average kinetic energy of the WIMP is:
 \beq \langle T\rangle \approx50~\mbox{keV}
\frac{m_{\chi}}{100~\mbox{GeV}}
 \label{kinen}
 \eeq
 So for sufficiently heavy WIMPs the
average energy may be adequate \cite{EFL88} to allow  scattering  to low lying excited states of of certain targets, e.g. of 
$57.7~$keV for the $7/2^{+}$ excited state of $^{127}$I, the 39.6 keV for the first excited $3/2^{+}$ of $^{129}$Xe, the 35.48 keV for the first excited $3/2^+$ state of $^{125}$Te  and the 9.4 keV for the first excited $7/2^{+}$ state of $^{83}$Kr .

The first requirement  is to have a large elementary nucleon cross section. In most models one can have both the scalar and spin induced cross section. In the case of the standard recoil experiments the coherent cross section is expected to dominate over the spin cross section especially for large $A$, unless the spin induced amplitude is very large at the quark level. This is possible in some models  like supersymmetry in the co-annihilation region \cite{Cannoni11}, where the ratio of the spin to coherent nucleon cross section, depending on $\tan{\beta}$ and the WIMP mass, which is in the range 200-500 GeV,  can be as large as $10^3$. In a region of the model space the ratio of the elastic spin cross section to the coherent can be as large as 10$\%$. More recent calculations in the supersymmetric $SO(10)$ model \cite{Gogoladze13}, also in the co-annihilation region, predict large spin to coherent cross section ratio, of the order of $2\times 10^{3}$  and a WIMP mass of about 850 GeV. Models of exotic WIMPs, like Majorana particles with  spin  3/2 \cite{SavVer13}, also can lead to large nucleon spin cross sections, which satisfy the relic abundance constrain.  These large nucleon cross sections may also allow inelastic scattering to the excited states with observable rates.
In fact we expect that the branching ratio to the excited state will be enhanced  in the presence of non zero energy threshold, since only the total rate to the ground state transition  will be affected and reduced, while this will have a negligible effect on the rate due to the inelastic scattering. The analysis is simplified if, as expected from  particle models as well as the structure of the nucleon,  the isovector nucleon cross section is dominant. 
\section{The spin dependent WIMP-Nucleus scattering}
The spin dependent WIMP-Nucleus cross section is typically expressed in terms of the WIMP-nucleon cross section, which contains the elementary particle parameters entering the problem at the quark level. For our purposes    it is adequate to have at the quark level two parameters, one isoscalar $\alpha_0(q)$  and one isovector  $\alpha_1(q)$. In going to the nucleon level one must   transform these  two amplitudes by suitable
renormalization factors
given in terms of the
quantities $\Delta q$ prescribed by Ellis \cite{JELLIS}, namely $\Delta
u=0.78 \pm 0.02$, $\Delta d=-0.48\pm 0.02 $ and $\Delta s=-0.15
\pm 0.02$, i.e.
\begin{align} &a_0=\alpha_0(q)\hspace{2pt}(\Delta u+\Delta
d+\Delta s)=0.13\hspace{2pt}\alpha_0(q), \nonumber \\
& a_1=\alpha_1(q)\hspace{2pt}(\Delta u-\Delta
d)=1.26\hspace{2pt} \alpha_1(q).
\end{align} 
In other words the isovector component
is renormalized as is expected for the axial current, while the
isoscalar component is suppressed,
 consistent with the EMC effect, i.e. the fact that only a tiny fraction of the spin of the nucleon is coming from the spin of the quarks.
   Thus in general,
barring very unusual circumstances at the quark level, the
isovector component is expected to be dominant. It is for this
reason that we started our discussion in the isospin basis.
From these amplitudes one  is able to calculate the isoscalar and isovector nucleon cross section. After that, within the context of a nuclear model, one can obtain the nuclear matrix element:
\beq
  |ME|^2 =a_1^2 S_{11}(u)+a_1 a_0 S_{01}(u)
  +a^2_{0}S_{00}(u),
\eeq
where $S_{ij}$ are the spin structure functions, which depend on the momentum transferred to the nucleus.  It  is useful to define the structure functions 
$$F_{ij}=\frac{S_{ij}}{\Omega_i \Omega_j}$$
with $\Omega_1,\,\Omega_0$ the isovector and isoscalar static spin matrix elements. Then the functions $F_{ij}$ take the value unity at momentum transfer zero. Furthermore these new  structure functions are approximately the same \cite{DIVA00,DivVer13}. Then to a good approximation
 \begin{align}
 |ME|^2 =\left (a_1^2  \Omega^2_1+a_1 a_0   \Omega_0 \Omega_1+a^2_{0} \Omega^2_0 \right) F_{11}(u).
  \label{Eq:MEsq}
  \end{align}
Thus, leaving aside the factor $F_{11}$, which will be taken care of explicitly in the expression for the event rate (see below),
   the  nuclear static spin cross section becomes
   \beq
  \sigma_A^{\mbox{\tiny{spin}}}=
  \frac{\left (  \Omega^2_1\sigma_1(N)+2\mbox{sign}(a_1 a_0)
   \Omega_0 \Omega_1\sqrt{\sigma_0(N)\sigma_1(N)}+ \Omega^2_0 \sigma_0(N)\right)}{3}.
  \label{Eq:nosf}
  \eeq
where $\sigma_0(N)$ and $ \sigma_1(N)$ are the elementary nucleon isoscalar and isovector  cross sections. Instead of these two  one could have written an equivalent expression in terms of the proton and neutron elementary cross 
sections \cite{DivVer13} $\sigma_p$ and $\sigma_n$.\\The above expressions look complicated. In most models, however, e.g. in the case  the WIMP is the Lightest Supersymmetric Particle 
(LSP),  one encounters $Z$ exchange and  $Z-qq$  coupling  are purely isovector. The same is true in models involving  heavy neutrinos as WIMPs, e.g.  Kaluza Klein theories in models with Universal Extra Dimensions \cite{OikVerMou}. In the latter case it is possible to have both an isoscalar and isovector component  if the WIMP happens to be a vector boson. Only isovector amplitudes are predicted also in the case of exotic spin 3/2 particles \cite{SavVer13}.  
 Since  the isoscalar nucleon cross section is expected to be negligible, the above expression reduces to 
	 \beq
  \sigma_A^{\mbox{\tiny{spin}}}=
  \frac{ \Omega^2_1\sigma_1(N)}{3}.
  \label{Eq:nosf0}
  \eeq
	
	Regarding the spin nucleon cross sections at this point there exist some experimental limits, namely for $^{129}$Xe and $^{131}$Xe \cite{Xenon100.11} and $^{19}$F \cite{COUPP12,SIMPLE12,PICASSO12,PICASSO09,PICASSO08}. The first group extracts a limit on the neutron cross section of $\sigma_n=2\times 10^{-40}$cm$^2=2\times 10^{-4}$pb, while from the latter  the limit $\sigma_p=1\times 10^{-28}$cm$^2=1.0\times 10^{-2}$pb. These limits were based on nuclear physics considerations, namely selecting the predominant nuclear spin matrix elements. Thus they cannot exclude the presence of the other elementary  nucleon cross section. On the other hand, if the elementary amplitude is purely isovector, these limits would imply  $\sigma_1=\sigma_p+\sigma_n=1.02\times 10^{-2}$pb.  We should mention in passing that the nuclear matrix elements for $^{19}$F are expected to be much more
	reliable \cite{DIVA00,DivVer13}.  Actually this problem will remain open until all the three nucleon cross sections (scalar, proton spin and neutron spin) can be determined along the lines previously suggested \cite{CannVerGom11}, after sufficient experimental information on at least three suitable  targets becomes available.
	Anyway in the present work we will not commit ourselves to any particular model, but for orientation purposes we will use the value \cite{SavVer13} of 
$\sigma_1(N)\approx 1.7 \times 10^{-38}$cm$^{-2}=1.7 \times 10^{-2}$pb.
\section{The structure of the  $^{83}$Kr nucleus}
 A complete calculation of the relevant spin stucture should be done
along the lines  previously done for other targets \cite{Ress}, 
\cite{PITTEL94}, \cite{VogEng}, \cite {IACHELLO91}, \cite{NIKOLAEV93},
\cite{SUHONEN03},\cite{Nsuhonen} and more recently \cite{MeGazSCH11},\cite{MeGazSCH12} in the  the shell model framework.  We will not concern ourselves with other simplified  models, e.g schemes of deformed rotational nuclei
 \cite{EngVog00}.

 A  summary of some nuclear ME involved in elastic and inelastic scattering can be found elsewhere \cite{SavVer13}, \cite{VerEjSav13}.

In principle one may have to go a step further in improving the structure functions, since it has recently been found that in the evaluation  of the ground state (gs) spin structure functions  some additional input is needed, namely:
\begin{itemize}
\item One has to consider the nucleon form factor \cite{DIVA00},\cite{DivVer13} entering the isovector axial current:
 \beq
 {\bf J}=\frac{g_A(q)}{g_A}\Sbf,\,\quad \Sbf=\bfs -\frac{\left ( \bfs.{\bf q}\right ){\bf q}}{q^2+m_{\pi}^2}
 \label{Eq:gAFf}
 \eeq
 If we choose as a $z$-axis the direction of momentum we find:
 \beq
 {\bf J}_m= \left \{ \begin{array}{l}\left (1-\frac{q^2}{q^2+m^2_{\pi}}\right )\bfs_m, \quad m=0\\ \bfs_{m}, \quad m=\pm1\end{array} \right.
 \eeq
Thus only the longitudinal component of the transition operator is
modified by the inclusion of the nucleon form factor. Thus, even
at sufficiently high momentum transfers, only 1/3 of the
differential rate will be affected.
\item One has to consider   effects arising from possible exchange currents.\\ 
Such currents lead to effective 2-body contributions obtained in the context of
  Chiral Effective Field Theory (EFT) \cite{MeGazSCH12}.
\end{itemize} 
We will elaborate further on this point,  however, since it has been shown that such effects are independent of the target nucleus \cite{DivVer13}. Thus they can be adequately contained in  a retardation  factor of the isovector amplitude, which  takes the value of 
about 0.8. Thus one can absorb such effects in the isovector nucleon spin cross section, which anyway  in practice it may have to be extracted from experiment, once the spin induced rates have been observed. We expect such a treatment will  also work in the structure functions involving transitions to excited states.
\section{Shell-Model Interpretations}\label{discuss_SM}
To interpret the experimental data, shell-model calculations have been carried out in the
28-50 valence shell composed of the $1p_{3/2}$, $0f_{5/2}$, $1p_{1/2}$, $0g_{9/2}$ orbitals.
The calculations have been performed with recently available effective interaction by Brown
and Lisetskiy \cite{jj44b} and shell model code NuShellX \cite{NuShellX}. The single-particle energies employed in
conjunction with the jj44b interaction are -9.6566, -9.2859, -8.2695, and -5.8944 MeV for
the $1p_{3/2}$, $0f_{5/2}$, $1p_{1/2}$, $0g_{9/2}$ orbitals, respectively. 
The ground state is correctly reproduce by present calculation.
The first excited $7/2^+$ state is predicted at 100 keV, while corresponding experimental value is 9 keV.

In Table \ref{t_be2}, we have compared experimental $B({\rm E}2)$ and $B({\rm M}1)$ values. The overall agreement
is reasonable, at least for B(E2) . The calculated $B({\rm E}2 : 7/2_1^+ \rightarrow 9/2_1^+)$ value is 11.62 W.u., while the corresponding
experimental value is 21.3(17) W.u. The structure of $9/2_1^+$ state is $\nu(g_{9/2})^{-3}$ (with probability $\sim$ 16\%).
The same configuration is predicted for the first excited state ($\sim$ 26\%). The calculated $B({\rm M}1 : 7/2_1^+ \rightarrow 9/2_1^+)$ value
is 0.0028 W.u., while the corresponding experimental value is 0.0095 W.u.  The calculated values of the quadrupole and magnetic moments are shown in Table
\ref{t_qm}.
These results are in reasonable agreement with the available experimental data. In particular, they reproduce correctly 
the sign of the quadrupole and magnetic moments. 

\begin{table}[h]
\caption{Calculated $B(E2)$ and $B(M1)$ values for
 $^{83}$Kr isotope with standard effective charges: e$_{\rm eff}^\pi$=1.5$e$, e$_{\rm eff}^\nu$=0.5$e$
 and for $g_{s}^{\rm eff}$ = $g_{s}^{\rm free}$
 (the experimental $\gamma$-ray energies corresponding to these transitions are also shown).}
\begin{tabular}{c|ccccc}
\hline
                                    &     & ~$B({\rm E}2) ({\rm W.u.})$ &      &~ $B({\rm M}1) ({\rm W.u.})$\\
\cline{3-4} \cline{5-6}
 $J_i^\pi \rightarrow J_f^\pi $       & $E_\gamma$ (keV) &~ Expt.   &~  jj44b  &~  Expt.   &~  jj44b\\
\hline
$7/2_1^+$ $\rightarrow$ ~$9/2_1^+$   &~~9.4       &~~ 21.3(17)      &~ 11.62   &~  0.0095      &~  0.0028  \\
\hline                   
\end{tabular}
\label{t_be2}
\end{table}
\begin{table}[h]
\caption{ Electric quadrupole moments, $Q_s$ (in $e$b) and magnetic
  moments, $\mu$ (in $\mu_N$),
 (the effective charges $e_p$=1.5, $e_n$=0.5  and for $g_{s}^{\rm eff}$ = $g_{s}^{\rm free}$are used in the calculation).}
\label{t_q}
\begin{center}
\begin{tabular}{c|c|c|c|c}
\hline
  $J^{\pi}$  &$Q_{s,{\rm exp}}$ &~$Q_{s,{\rm jj44b}}$    &$\mu_{\rm exp}$ &~$\mu_{\rm jj44b}$\\		   
\hline
             9/2$^+$    &~~ +0.26 (3)   &~~+0.34   &~~ -0.970669 (3)   &-1.412 \\
             7/2$^+$     &~~ +0.495 (10)   & ~~ +0.41   &~~ -0.943 (2)     & -1.099          \\ 
\hline                             
\end{tabular}
\end{center}
\label{t_qm}
\end{table} 

\begin{figure}
\begin{center}
\subfloat[]
{
\includegraphics[width=.5\textwidth]{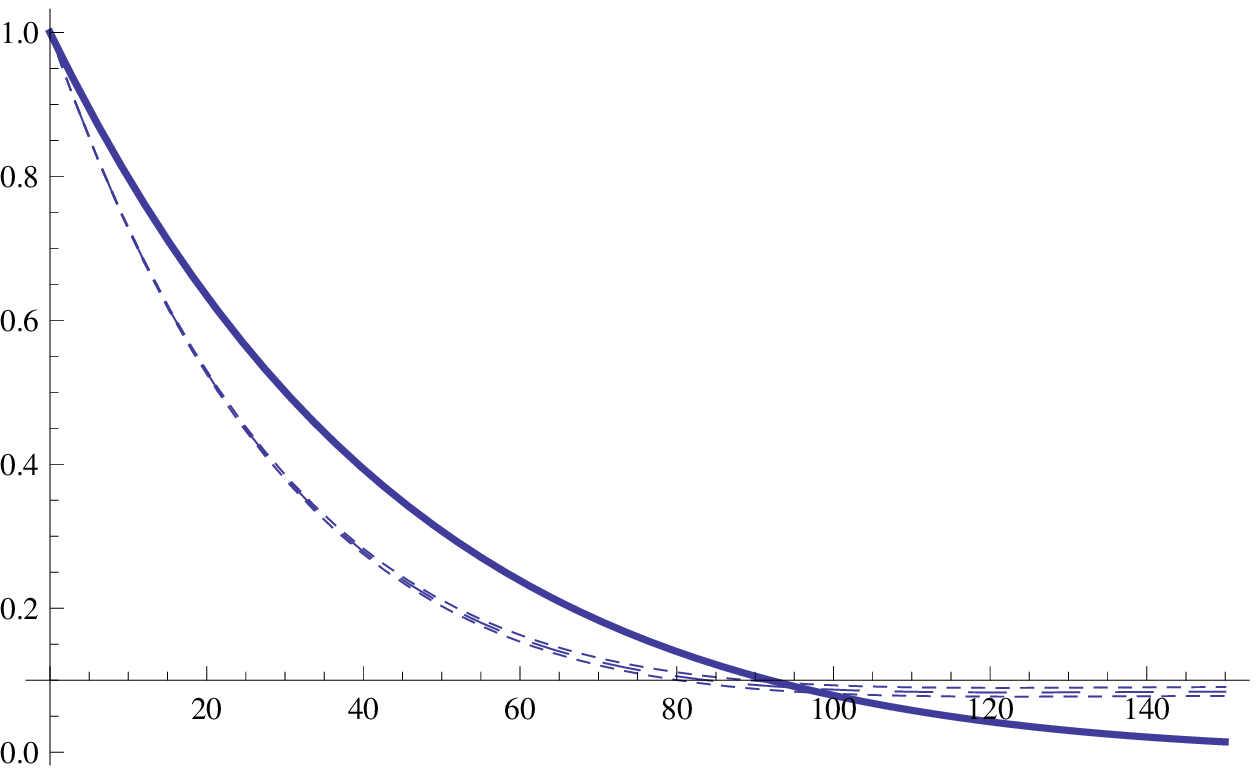}
}
\subfloat[]
{
\includegraphics[width=.5\textwidth]{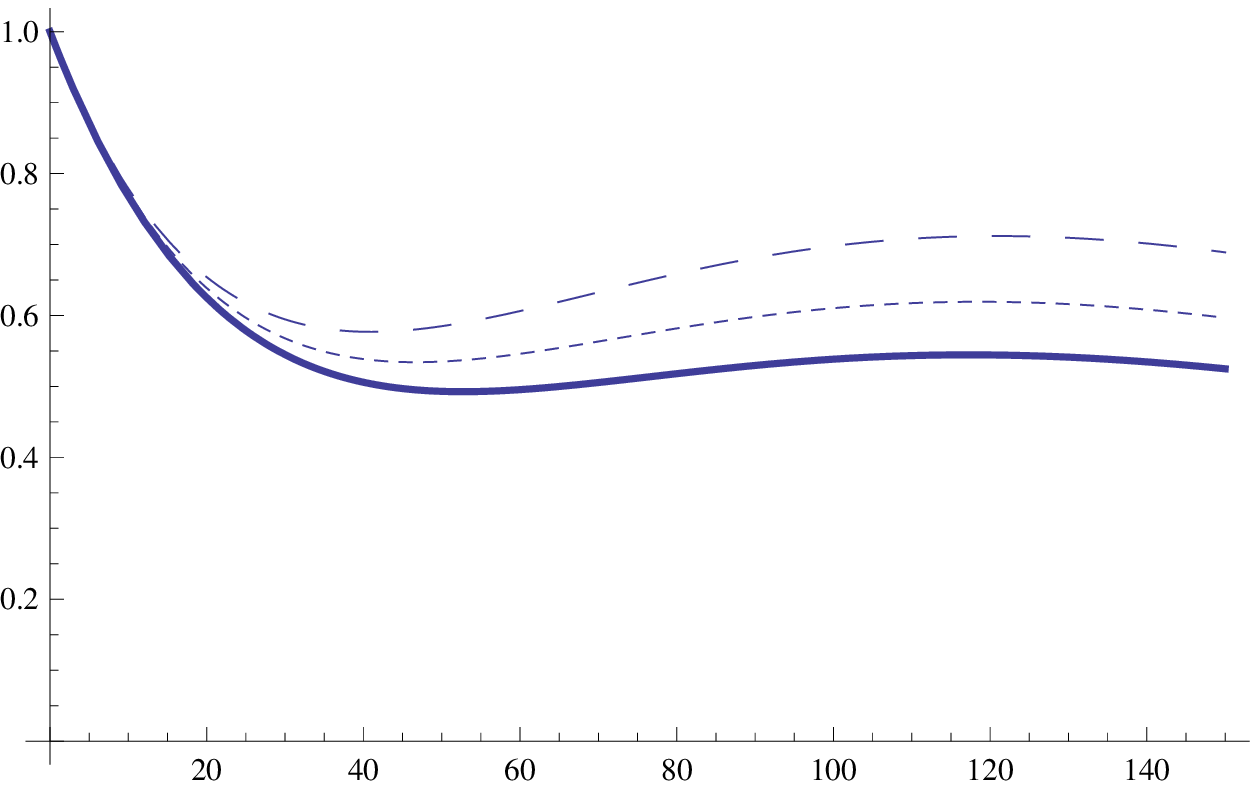}
}
\\
{\hspace{-1.5cm} $E_R\rightarrow$keV}

\caption{Elastic spin  structure function $F^{e\ell}_{ij} ( E_R)$ for the isospin modes $i,j={11},{01},{00}$ as functions of the energy transfer in keV in the  case of the target $^{83}$Kr (a). For comparison we present the square of the form factor entering the coherent mode (solid line). We also show the inelastic spin structure function $F^{in}_{ij}(E_R)$ (b) . Note that the elastic structure functions, when normalized to unity at momentum transfer zero, are essentially independent of isospin. In the case of the inelastic scattering the structure functions are not very much depressed as function of the energy transfer, compared to those of the elastic scattering.
The solid line, dotted line and dashed line  correspond to $F_{11}, F_{01}$ and $F_{00}$ respectively.
 \label{fig:FFspin}}
\end{center}
\end{figure}

 The thus obtained static spin MEs are
$$ \Omega_0=1.037,\,\Omega_1= -1.018  \mbox{ for elastic transitions},$$
$$ \Omega_0= -4.880\times 10^{-2},\,\Omega_1= 4.421\times 10^{-2} \mbox{ for inelastic transitions}.$$
Looking at these matrix elements we notice the following:
\begin{itemize}
\item The nuclear wave functions do not favor the isoscalar contribution to overcome the expected suppression  of the corresponding  isoscalar amplitude in going from the quark to the nucleon level discussed above.
\item The inelastic transition is not predominantly  a Gamow-Teller-like transition (that is, driven by the $\pmb{\sigma}$ operator), but but it is dominated by higher multipoles.\\
 The disadvantage of small value of the spin ME in the inelastic transition may, however, be partly overcome by the fact that, unlike the EM decay of the excited state, a relatively high momentum can be transferred to the nucleus by the WIMP and the excited state structure  function does not fall fast with the transferred momentum.
\end{itemize}
The structure functions relevant to our calculation are exhibited in Fig.  \ref{fig:FFspin}.

\section{The formalism for the WIMP-nucleus differential event rate}
The expression for the elastic differential event rate is well known, see e.g. \cite{SavVer13},\cite{DivVer13} . For the reader's convenience and to establish our notation we will briefly present the essential ingredients here. We will begin with the more familiar time averaged coherent rate, which   can be cast in the form:
\beq
\left .\frac{d R_0}{ dE_R }\right |_A=\frac{\rho_{\chi}}{m_{\chi}}\,\frac{m_t}{A m_p}\, \left (\frac{\mu_r}{\mu_p} \right )^2\, \sqrt{<\upsilon^2>} \,\frac{1}{Q_0(A)} A^2 \sigma_N^{\mbox{\tiny{coh}}}\left .\left (\frac{d t}{du}\right ) \right |_{\mbox{ \tiny coh}},\, \left .\left (\frac{d t}{d u}\right )\,\right |_{\mbox{\tiny coh}}=\sqrt{\frac{2}{3}}\, a^2 \,F^2(u)  \, \Psi_0(a \sqrt{u})
\label{drdu}
\eeq
with with $\mu_r$ ($\mu_p$) the WIMP-nucleus (nucleon) reduced mass and $A$ is the nuclear mass number. $ m_{\chi}$ is the WIMP mass, $\rho(\chi)$ is the WIMP density in our vicinity, assumed to be 0.3 GeV cm$^{-3}$,  and $m_t$ the mass of the target. $u$ is the recoil energy  $E_R$ in dimensionless units introduced here for convenience, $u=\frac{1}{2}(q b)^2=A m_p E_R$, with $A$ the nuclear mass number of the target and $b$ the nuclear harmonic oscillator size parameter characterizing the nuclear wave function. It simplifies the expressions for the nuclear form factor and structure functions. In fact:
\begin{equation}
 u=\frac{E_R}{Q_0(A)}~~,~~Q_{0}(A)=[m_pAb^2]^{-1}=40A^{-4/3}\mbox{ MeV}
\label{defineu}
\end{equation}

The factor $\sqrt{2/3}$ in the above expression is  $\upsilon_0/\sqrt{\langle \upsilon ^2\rangle}$ since in Eq. (\ref{drdu}) the WIMP flux is given in units of $\sqrt{\langle \upsilon ^2\rangle}$ appears. In the above expressions  $a=(\sqrt{2} \mu_r b \upsilon_0)^{-1}$, $\upsilon_0$ the velocity of the sun around the center of the galaxy and 
 $F(u)$ is the nuclear form factor. Note that the parameter $a$ depends both on the WIMP mass, the target and the velocity distribution. 

For the axial current (spin induced) contribution
 one finds for the elastic WIMP-nucleus scattering:
\beq
\left .\frac{d R_0}{ dE_R }\right |_A=\frac{\rho_{\chi}}{m_{\chi}}\,\frac{m_t}{A m_p}\, \left (\frac{\mu_r}{\mu_p} \right )^2\, \sqrt{<\upsilon^2>} \,\frac{1}{Q_0(A)} \, \sigma_A^{\mbox{\tiny{spin}}}\left .\left (\frac{d t}{du}\right ) \right |_{\mbox{ \tiny spin}},\,\left . \left (\frac{d t}{d u}\right )\right |_{\mbox{\tiny spin}}=\sqrt{\frac{2}{3}} \, a^2  \, F_{11}(u)   \,  \Psi_0(a \sqrt{u})
\label{drdus}
\eeq
where $F_{11}$ is the isovector spin response function, i.e. $F_{e\ell}$ for the elastic case and $F_{in}$ for the inelastic one. Their behavior is exhibited in Fig. \ref{fig:FFspin}. 
\\We notice that the only difference in the shape of the spectrum between the coherent and the spin comes from nuclear physics. Since the square of the form factor and the spin structure functions have approximately similar shapes, we expect the corresponding differential shapes to be the same, and only the scale to be different.

The function $\Psi_0(x) $  is defined by $\Psi_0(x)=g(\upsilon_{min},\upsilon_E(\alpha))/\upsilon_0$, where $\upsilon_0$ is the velocity of the sun around the center of the galaxy.
   $g(\upsilon_{min},\upsilon_E(\alpha))$   
 depends on the velocity distribution in the local frame through  the minimum WIMP velocity for a given energy transfer, i.e.
\beq \upsilon_{min}=\sqrt{\frac{A \,m_p \,E_R}{2 \, \mu^2_r}}. \eeq

The above way of writing the differential event rates we have explicitly separated the three important factors:
\begin{itemize}
\item the kinematics,
\item the nuclear cross section $A^2 \sigma_N$ or $\sigma_A^{\mbox{\tiny{spin}}}$
\item the combined effect of the folding of the velocity distribution and  the form factor or the nuclear structure function.
\end{itemize}
 
For the Maxwell-Boltzmann (M-B ) distribution in the local frame it is defined as
follows:
\begin{align}
&g(\upsilon_{min},\upsilon_E)=\frac{1}{\big
(\sqrt{\pi}\upsilon_0 \big
)^3}\int_{\upsilon_{min}}^{\upsilon_{max}}e^{-(\upsilon^2+2 \vbf .
\vbf_E+\upsilon_E^2)/\upsilon^2_0} \, \upsilon  \,
d\upsilon  \, d \Omega, \nonumber
\\
& \upsilon_{max}=\upsilon_{esc},
\end{align}
where $\vbf_E$ is the velocity of the Earth, including the velocity of the sun
around the galaxy. We have neglected the velocity of the Earth around the sun, since we ignore the time dependence (modulation) of the rates.
The above upper cut-off value in the M-B is usually put in by hand.
 Such a cut-off comes in naturally,
however, in the case of velocity distributions obtained from the halo
WIMP mass density in the Eddington approach \cite{VEROW06}, which,
in certain models, resemble a M-B distribution \cite{JDV09}.


Integrating the above differential rates we obtain the total rate including the time averaged rate 
for each mode  given by:
\beq
R_{\mbox{\tiny coh}}=\frac{\rho_{\chi}}{m_{\chi}} \, \frac{m_t}{A m_p}  \left ( \frac{\mu_r}{\mu_p} \right )^2  \sqrt{<\upsilon^2>} \, A^2 \, \sigma_N^{\mbox{\tiny{coh}}} \, t_{\mbox{\tiny coh}},\quad t_{\mbox{\tiny coh}}=\int_{E_{th}/Q_0(A)}^{(y_{\mbox{\tiny esc}}/a)^2} \, \left .\frac{dt}{du}\right |_{\mbox{\tiny coh}}du,
\label{Eq:Trates}
\eeq
\beq
R_{\mbox{\tiny spin}}=\frac{\rho_{\chi}}{m_{\chi}} \, \frac{m_t}{A m_p} \, \left ( \frac{\mu_r}{\mu_p} \right )^2  \sqrt{<\upsilon^2>} \, \sigma_A^{\mbox{\tiny{spin}}} \, t_{\mbox{\tiny spin}}, \quad t_{\mbox{\tiny spin}}=\int_{E_{th}/Q_0(A)}^{(y_{\mbox{\tiny esc}}/a)^2} \, \left .\frac{dt}{du}\right |_{\mbox{\tiny spin}}du
\label{Eq:Tratec}
\eeq
for each mode (spin and coherent). $E_{th}(A)$ is the energy threshold imposed by the detector.

These expressions contain the following parts: i) The gross properties and kinematics ii) The parameter $t$, which contains the effect of the velocity distribution and the nuclear form factors iii) The WIMP-nuclear cross sections $A^2 \sigma_N$ or $\sigma_A^{\mbox{\tiny{spin}}}$. The latter, contains the nuclear static spin ME.  From the latter the elementary nucleon cross sections can be obtained, if one mode becomes dominant as already mentioned above. Anyway using the values for nucleon cross sections, $\sigma_N^{\mbox{\tiny{coh}}}$ in Eq. (\ref{Eq:Trates}) and $ \sigma_A^{\mbox{\tiny{spin}}}$ in Eq. (\ref{Eq:Tratec}), we can obtain the total rates. \\
Conversely, if only one mode is dominant, one can extract from the data the relevant nucleon cross section (coherent, spin isoscalar or spin isovector) or obtain exclusion plots on them. 

 \section{transitions to excited states}
Transitions to excited states are normally energetically suppressed, except in some odd nuclei that have low lying excited states. Then spin mediated transitions to excited states are possible. The most favorable are expected to be those that, based on the total angular momentum and parity of the states involved, appear to be  Gamow-Teller like transitions. A good possibility is  the $7/2^+$ state of $^{83}$Kr,  which is at $9.5$ keV above the $9/2^+$ ground state.
\subsection{Isotope considerations}
Transitions to the first excited states by spin interacting WIMPs (SD WIMPs) can occur in the case of some odd A targets, if the relevant excitation energy is $E \le 100 $ keV. This is due to the high velocity tail of the M-B distribution, so that a reasonable amount of the WIMP energy may be transferred to the recoiling nucleus.

Possible odd mass nuclei involved in DM detectors used for WIMP searches are the 57.6 keV state in $^{127}$I and the 39.6 keV state in $^{129}$Xe. The spin excitations of these states are not favored, because the dominant components of the relevant wave functions  are characterized by $\Delta \ell\ne 0$, i.e. $\ell$ forbidden transitions.  Nevertheless the spin transitions are possible, due to the small components  as seen from the M1 $\gamma $ transition rates. Other possibilities are, of course, of interest, e.g. 
$^{83}$Kr is a good experimental candidate. One can easily enrich it isotopically
and it can make a good liquid or gas detector. 
 Thus it is quite realistic to study the inelastic excitations in this nucleus to search for SD WIMPs. 

In fact the experimental observation of the inelastic excitation has several advantages. The experimental feasibility is discussed in section VIII.

\begin{table} [h]
\caption{Inelastic spin excitations of $^{83}$Kr. $A$ : natural abundance ratio, $E$ : excitation energy, $J_i$: ground state spin parity, $J_f$ : excited state spin parity, and  $T_{1/2 } $: half life. For comparison we present the same quantities for  $^{125}$Te.}
\label{tab:ej1}
\begin{center}
\begin{tabular}{cccccc} 
  \\ 
\hline
Isotope &   $A(\%)$ &   $E$ (keV) &   $J_i$ &   $J_f$ &  $T_{1/2} $(ns) \\        
\hline
$^{83}$Kr    &       11.5  &        9.5  &    9/2$^+$  &  7/2$^+$ &    154\\  
$^{125}$Te &       7.07 &      35.5  &    1/2$^+$  &   3/2$^+$ & 1.48  \\     

 \hline
\end{tabular}
\end{center}
\medskip 
\end{table}  

\subsection{Kinematics}

\begin{figure}
\begin{center}
\subfloat[]
{
\rotatebox{90}{\hspace{0.0cm} $\left (E_R\right)_{min} \rightarrow$keV}
\includegraphics[height=0.5\textwidth]{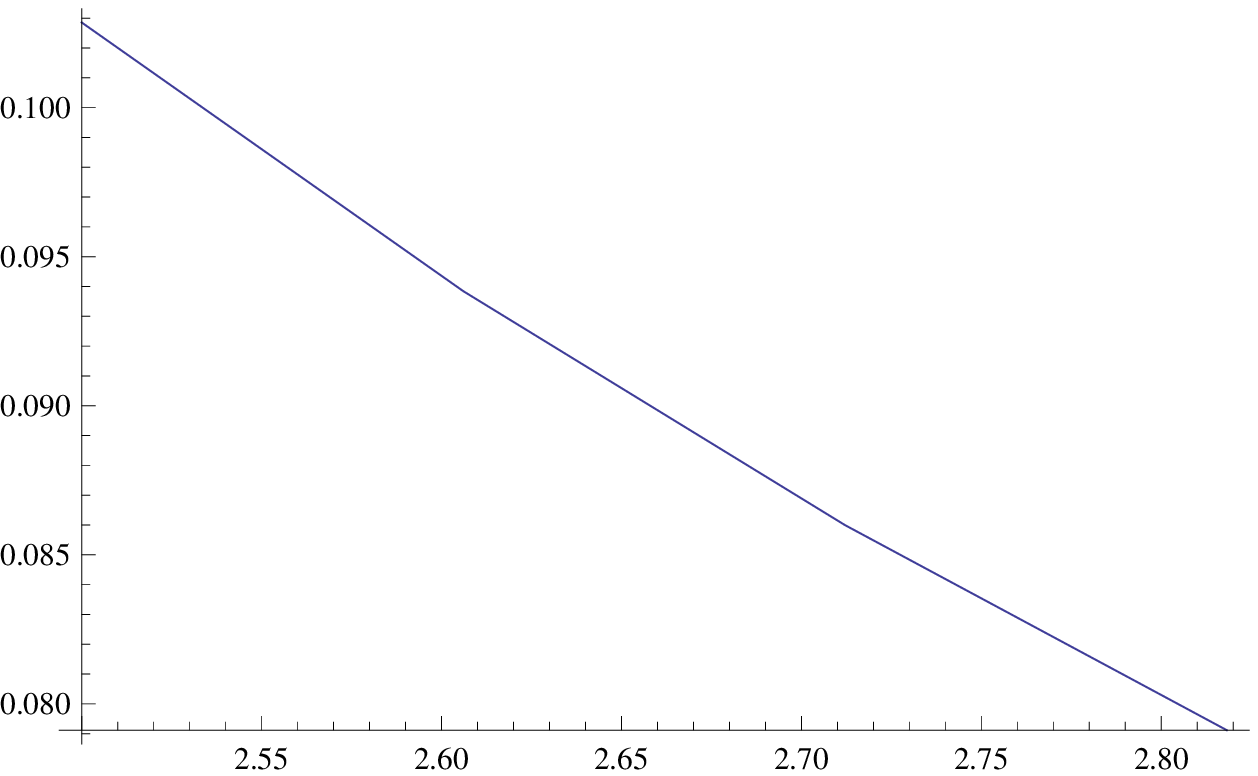}
}\\
\subfloat[]
{
\rotatebox{90}{\hspace{0.0cm} $\left (E_R\right)_{min} \rightarrow$keV}
\includegraphics[height=0.5\textwidth]{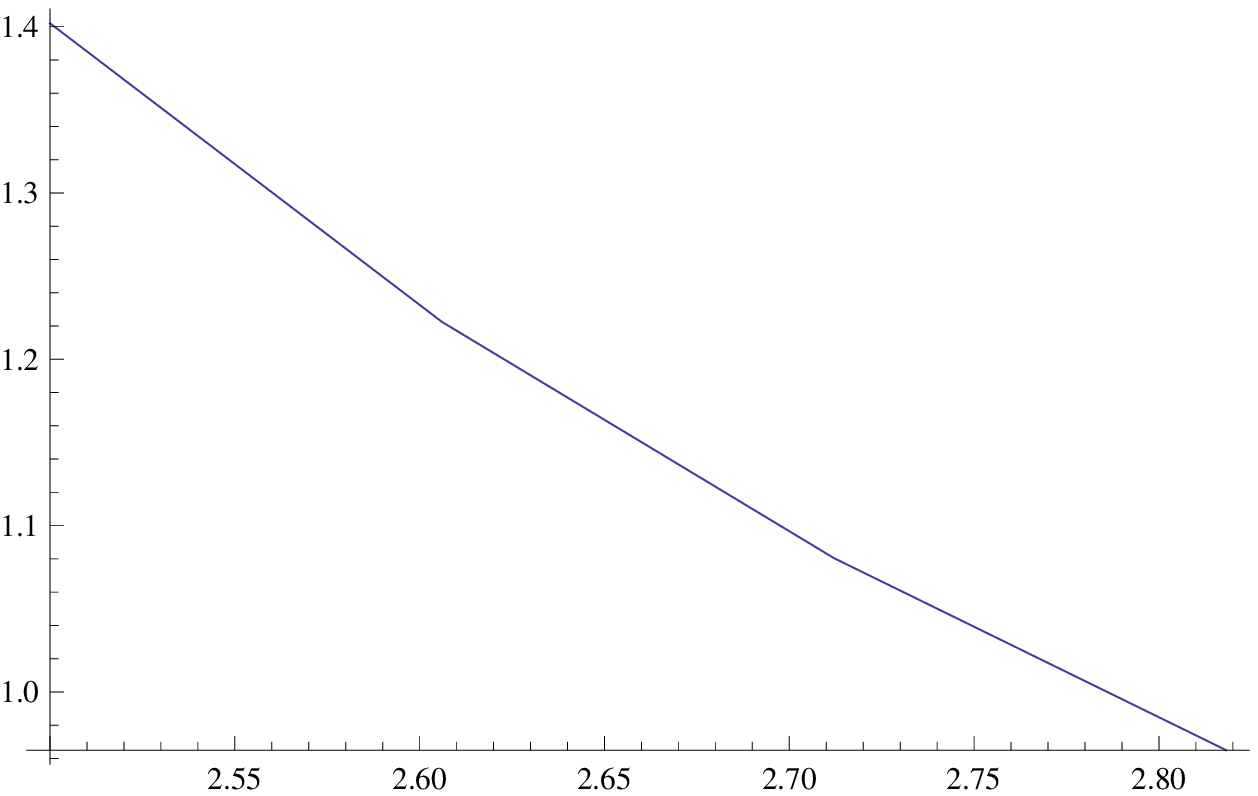}
}
\\
{\hspace{-2.0cm} $y_{esc}=\upsilon_{esc}/\upsilon_0 \rightarrow$}\\
\caption{ The minimum required energy transfer as a function of the escape velocity for $^{83}$Kr (a). For comparison we present the same quantity for  $^{125}$Te (b).
 \label{fig:fumin}}
\end{center}
\end{figure}

The evaluation of the differential rate for the inelastic transition proceeds in a fashion similar to that of the elastic case discussed above, except:
\begin{enumerate}
\item The transition spin matrix element must be used.
\item The transition spin response function must be used.
For Gammow-Teller like transitions, it does not vanish for zero energy transfer. So it can be normalized to unity, if the static spin value is taken out of the ME.
\item The kinematics is modified.
 The energy-momentum conservation reads:
 \beq
 \frac{-q^2}{2 \mu_r}+\upsilon \xi q -E_x =0, \quad E_x= \mbox {excitation energy}\Leftrightarrow -\frac{m_A}{\mu_r}E_R +\upsilon \xi \sqrt{2 m_A E_R}-E_x =0
 \eeq
 Clearly $\xi>0$ as before. Then $\xi<1$  and the reality of $E_R$ impose the condition:
 \beq
  \upsilon>\frac{E_x+\frac{m_A}{\mu_r}E_R}{\sqrt{2 m_A E_R}}
 \eeq
We thus find that for a given energy transfer $E_R$ the minimum allowed WIMP velocity is given by:
\beq
\upsilon_{min}=\frac{E_x+\frac{m_A}{\mu_r}E_R}{\sqrt{2 m_A E_R}}
\eeq
while the maximum and minimum energy transfers are limited by the escape velocity in the WIMP velocity distribution. We find that:
\beq
(E_R)_{min}\leq E_R\leq (E_R)_{max}
\eeq
with
\beq
\left (E_R\right) _{min}=\frac{\mu_r^2}{m_A}\left( \upsilon_{esc}^2-\frac{E_x}{\mu_r}-\sqrt{\upsilon_{esc}^4-2 \upsilon_{esc}^2\frac{E_x}{\mu_r}}\right),\,\left (E_R\right) _{max}=\frac{\mu_r^2}{m_A}\left( \upsilon_{esc}^2-\frac{E_x}{\mu_r}+\sqrt{\upsilon_{esc}^4-2 \upsilon_{esc}^2\frac{E_x}{\mu_r}}\right)
\eeq
In the case of elastic scattering  we recover the familiar formulas:
$$(E_R)_{min}=0,\,(E_R)_{max}=2 \frac{\mu_r^2}{m_A}\upsilon_{esc}^2$$
 From the above expressions it is clear that for a given nucleus and excitation energy only WIMPs with a mass above a certain limit are capable of causing the inelastic transition, i.e. 
\beq
m_{\chi}\geq m_A\left (\frac{1}{2}\upsilon^2_{esc}\frac{m_A}{E_x}-1\right )^{-1}\rightarrow
\eeq
$$  \left (m_{\chi}\right )_{min}=(4.6,\,19,\,34,\,21)\mbox{ GeV for } ^{83}\mbox{Kr},\,^{125}\mbox{Te},\,^{127}\mbox{I},\,^{129}\mbox{Xe  respectively}$$

 We find it simpler to  deal with the phase space in dimensionless units. Noticing that $u=(1/2) q^2 b^2$ and
 \beq
 \delta \left (\frac{-q^2}{2 \mu_r}+\upsilon \xi q -E_x \right )=\delta \left (-\frac{u}{\mu_r b^2}+\upsilon \xi\frac{\sqrt{2 u}}{b}-E_x \right )\Leftrightarrow \frac{b}{\upsilon \sqrt{2u}}\delta \left (\xi-\frac{E_x+u/(\mu_r b^2)}{\upsilon \sqrt{2u}}\right )
 \eeq
 we find:
 \beq
 \int q^2 d \xi dq \delta \left (\frac{-q^2}{2 \mu_r}+\upsilon \xi q -E_x \right )=\frac{1}{b^2 \upsilon} du
 \eeq
 i.e. we recover the same expression as in the case of ground state transitions.\\
 The above constraints now read:
 \beq
 y>a \frac{u+u_0}{\sqrt{u}}
 \eeq
 \beq
 u_0=\mu_r E_x b^2,\quad a=\frac{1}{\sqrt{2}\mu_r \upsilon_0 b},\quad u=\frac{E_R}{Q_0(A)}.
 \eeq
It should be stressed that for transitions to excited states the energy of recoiling nucleus must be above a minimum energy, which depends on the escape velocity, the excitation energy and the mass of the nucleus as well as the WIMP mass. This limits the inelastic scattering only for recoiling  energies  above  the values $\left(E_R\right)_{min}$.
 The minimum and maximum energy that can be transferred is:
 \beq
 u_{min}=\frac{1}{4}\left (\frac{y_{esc}}{a}-\sqrt{\left (\frac{y_{esc}}{a} \right )^2-4 u_0} \right )^2,\,
 u_{max}=\frac{1}{4}\left (\frac{y_{esc}}{a}+\sqrt{\left (\frac{y_{esc}}{a} \right )^2-4 u_0} \right )^2
 \eeq
 \end{enumerate}
 The maximum energy transfers $ u_{max}$  depend on the escape velocity $\upsilon_{exc}=620$km/s. Here we have denoted   $\upsilon_{exc}=y_{esc}\upsilon_0$, where $\upsilon_0$ is the characteristic velocity of the M-B distribution taken to be 220 km/s, i.e. $y_{esc}\approx2.84$. 
The values $u_{min}$, $u_{max}$, $\left(E_R\right)_{min}$ and $\left(E_R\right)_{max}$ relevant for the inelastic scattering of $^{83}$Kr and $^{125}$Te are shown in Table \ref{tab:tab2}.  The minimum required and the maximum allowed energy transfer depend on the WIMP escape velocity. The dependence of the maximum is not crucial, since the contribution to the total rate becomes negligible at high energy transfers. The behavior of the minimum in the range of the existing limits of the escape velocities \cite{LisSper11} is exhibited in Fig. \ref{fig:fumin}
 
\begin{table}[h]
\caption{ The kinematical parameters entering the inelastic scattering to the first excited 
state of $^{83}$Kr. For comparison we present the same quantities for  $^{125}$Te.}
 \label{tab:tab2}
\begin{center}
\begin{tabular}{|l|c|cccccr|} 
\hline\hline
target    & parameter                            &       & $m_{\chi}$(GeV)& & &   &     \\
\hline
          &                                     &20 & 50    & 100   & 200   & 500   & 1000 \\
\hline
$^{83}$Kr & $u_{\mbox{min}}/10^{-3}$              & 0.671   & 0.645&0.638&0.633&0.632&0.631\\
  & $u_{\mbox{max}}$                 &0.440  & 1.71&3.67&6.17&9.12&10.5 \\
          & $\left(E_R\right)_{\mbox{min}}$(keV) &0.078& 0.049& 0.049& 0.049& 0.048& 0.048 \\
          & $\left(E_R\right)_{\mbox{max}}$(keV) &51.6& 199&425&716&1059&1228 \\
\hline
$^{125}$Te & $u_{\mbox{min}}$                           &0.014    & 0.012& 0.012& 0.011& 0.011& 0.0.11 \\
           & $u_{\mbox{max}}$                 &0.45  & 2.35&8.95&11.69&20.00&24.86 \\
           &$\left(E_R\right)_{\mbox{min}}$(keV)&0.94 &0.78& 0.74& 0.72& 0.71& 0.71  \\
           &$\left(E_R\right)_{\mbox{max}}$(keV) &30.2&  157&400&786&1347&1670\\
\hline
\hline
\end{tabular}
\end{center}
\end{table}

 To avoid uncertainties arising from the relevant particle model, we will present the  rate to the excited relative to that to the ground state (branching ratio).
 The differential event rate for inelastic scattering takes a form similar to the one given by Eq. \ref{drdus} except that 
 $$\Omega_1\rightarrow \Omega_1^{\mbox{\tiny{inelastic}}},\, F_{11}(u)\rightarrow F_{11}(u)^{\mbox{\tiny{inelastic}}},\quad\Psi_0(a\sqrt{u})\rightarrow \Psi_0\left(a\frac{ u+u_0}{\sqrt{u}}\right ).$$
\section{Some results}
For purposes of illustration we will employ the  nucleon cross section of $1.7\times 10^{-2}$pb  obtained in a recent 
work \cite{SavVer13}, without committing ourselves to this or any other  particular model. Another input is the WIMP density in our vicinity, which will be taken to be 0.3 GeV cm$^{-3}$. Finally the velocity distribution with respect to the galactic center will be assumed to be a  M-B with a characteristic velocity $\upsilon_0=220$km/s and an upper cut off (escape velocity) of $2.84 \upsilon_0$.
 \subsection{The differential event rates} 
\begin{figure}
\begin{center}
\subfloat[]
{
\rotatebox{90}{\hspace{0.0cm} $\left .\left (\frac{dR_0}{dE_R}\right )\right|_A \rightarrow$events/(kg-y)/ keV}
\includegraphics[width=0.45\textwidth]{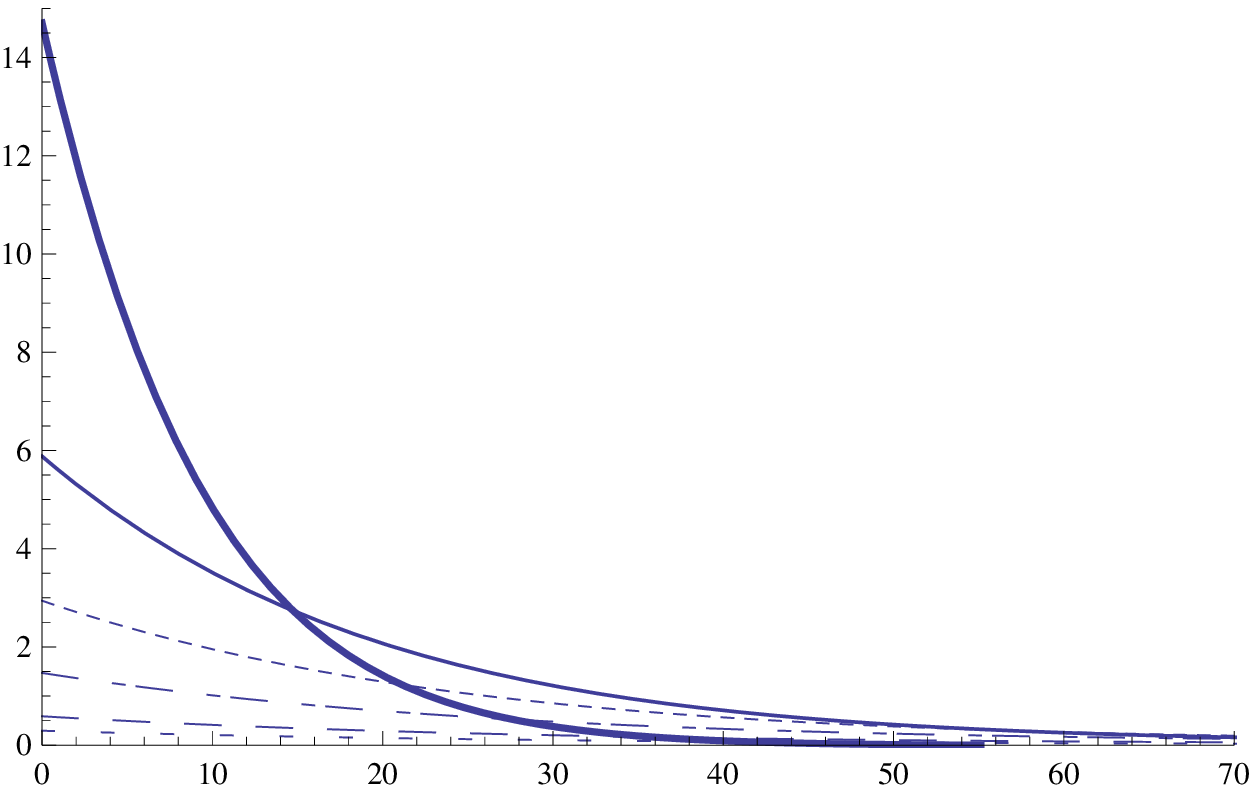}
}
\subfloat[]
{
\rotatebox{90}{\hspace{0.0cm} $\left .\left (\frac{dR_0}{dE_R}\right )\right|_A \rightarrow$events/(kg-y)/ keV}
\includegraphics[width=0.45\textwidth]{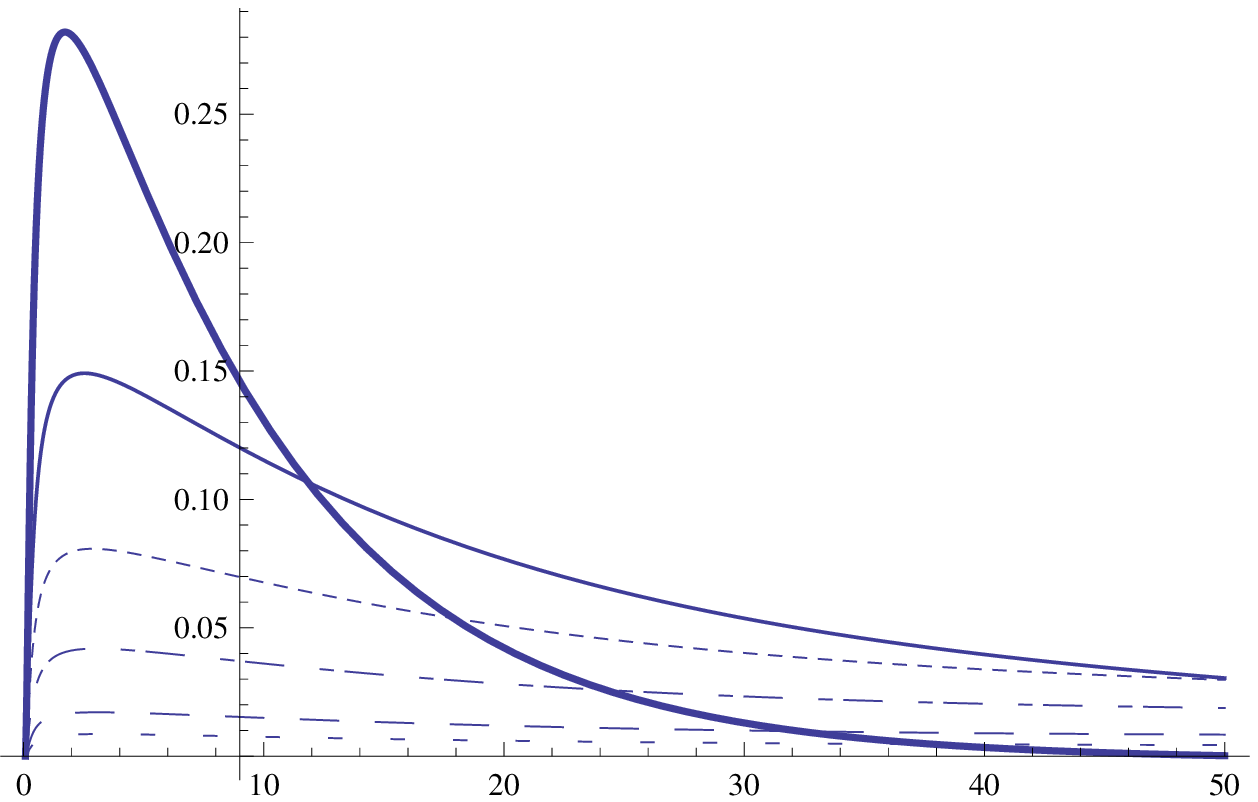}
}
\\
{\hspace{-2.0cm} $E_{R}\rightarrow$keV}
\caption{ Energy spectrum  for WIMP $^{83}$Kr elastic scattering (a) and that for the 9.4 keV excited state inelastic scattering (b). The thick solid, solid, dotted, dashed-dotted and dashed curves correspond to WIMP masses 50, 100, 200, 500 and 1000 GeV. }
\label{fig:dRdQKr}
\end{center}
\end{figure}

The differential event rates, perhaps the most  interesting from an experimental point of view,  depend on the WIMP mass, but we can only present them for some select masses.
Our results for the  elastic differential rates for typical WIMP masses are  exhibited in Fig. \ref{fig:dRdQKr}a. For comparison we present the differential event rates for transition to the excited state in Fig. \ref{fig:dRdQKr}b.

\begin{figure}
\begin{center}
\subfloat[]
{
\rotatebox{90}{\hspace{0.0cm} $\frac{dR(\mbox{{\small excited}})}{dE_R}/\frac{dR(\mbox{{\small gs}})}{dE_R}\times 10^{5}$}
\includegraphics[width=.4\textwidth]{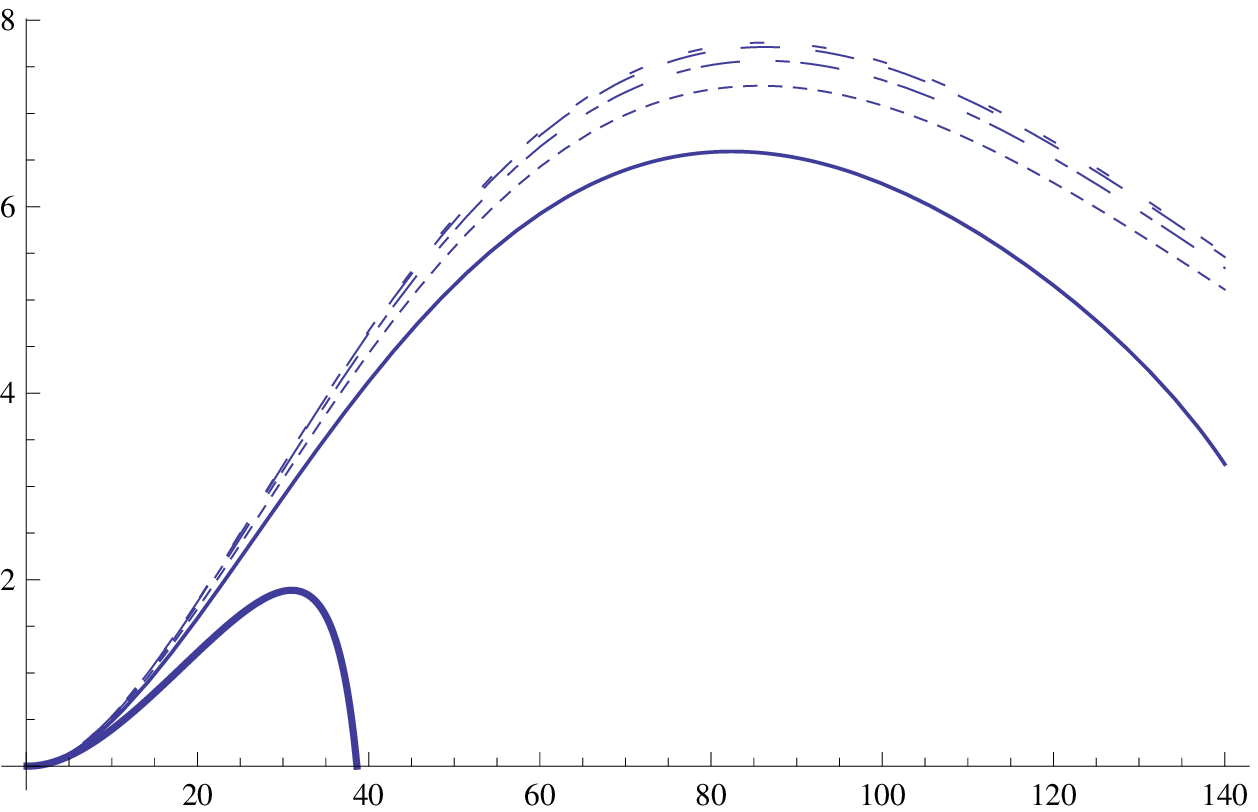}
}
\subfloat[]
{
\rotatebox{90}{\hspace{0.0cm} $\frac{dR(\mbox{{\small excited}})}{dE_R}/\frac{dR(\mbox{{\small gs}})}{dE_R}\times  10^{5}$}

\includegraphics[width=.4\textwidth]{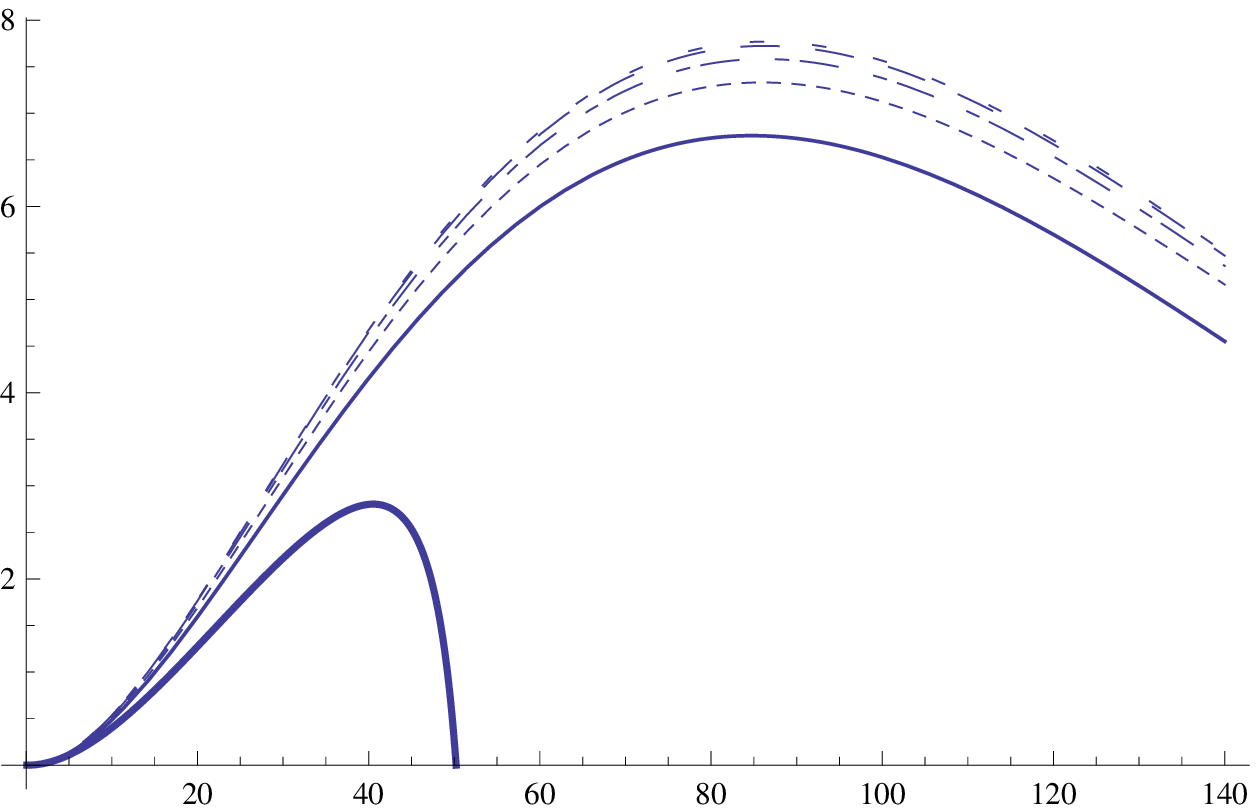}
}\\
{\hspace{-2.0cm} $E_R\rightarrow$keV}\\
\caption{ The ratio of the differential scattering rates, $\frac{dR(\mbox{{\small excited}})}{dE_R}/\frac{dR(\mbox{{\small gs}})}{dE_R}$ in units of $10^{-5}$, as a function of the recoil energy $E_R$ in keV in the case of the target $^{83}$Kr. The thick solid, solid, dotted, dashed-dotted and dashed curves correspond to WIMP masses 50, 100, 200, 500 and 1000 GeV. In panel (a) we show the results  for $y_{esc}=2.5$ and in (b) for $y_{esc}=2.8$.  The  range of the allowed  energy transfer, depends, of course, on the WIMP mass.
 The dependence on the escape velocity in the range of the accepted values is mild.
 \label{fig:difRex}}
\end{center}
\end{figure}

%
 We will  evaluate the branching ratio of the inelastic differential cross section to that of the ground state ignoring the coherent mode.
  We will restrict ourselves in the isovector transition. This is reasonable,  since  as we have already mentioned  the isoscalar is absent in most particle models and even if it appears at the quark level it  is expected to be suppressed  \cite{JELLIS93a} due to considerations related to the spin of the nucleon.  In such a case the ratio becomes independent of the elementary nucleon cross section. The obtained results are exhibited in Fig. \ref{fig:difRex}.
\subsection{Total rates}
\begin{figure}
\begin{center}
%
\includegraphics[width=0.8\textwidth]{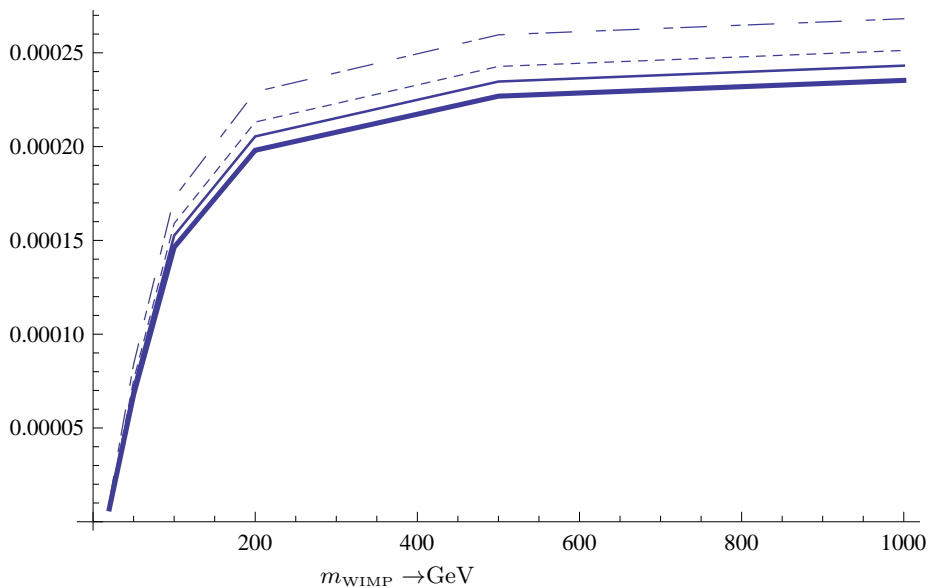}
\\
{\hspace{-2.0cm} $m_{\mbox{{\tiny WIMP}}}\rightarrow$GeV}
\caption{ The  ratio of the total scattering rates, $R(\mbox{excited})/R(\mbox{gs})$, as a function of the WIMP mass in GEV for $y=2.84 $ in the case  $^{83}$Kr. 
From bottom up the threshold values are $0,1,2,3,4,7$ and 10 keV. Only the spin mode has been taken into account. No quenching factor has   been used.
 \label{fig:RexcTh}}
\end{center}
\end{figure}

From expressions (\ref{Eq:Trates}) and (\ref{Eq:Tratec}), we can obtain the total rates.
It is instructive to start with a consideration of  the branching ratio of the total rates, which for the reasons discussed above is going to be independent of the elementary nucleon cross section.
Since the elastic scattering  event rate is reduced by the threshold effects, but the inelastic scattering is not affected by such effects, we expect the branching ratio to be increasing as the threshold energy is increasing. The situation is exhibited in Fig. \ref{fig:RexcTh}.

\begin{figure}
\begin{center}
\subfloat[]
{
\rotatebox{90}{\hspace{0.0cm} $\left . R_0\right|_A \rightarrow$events/(kg-y)}
\includegraphics[width=.45\textwidth]{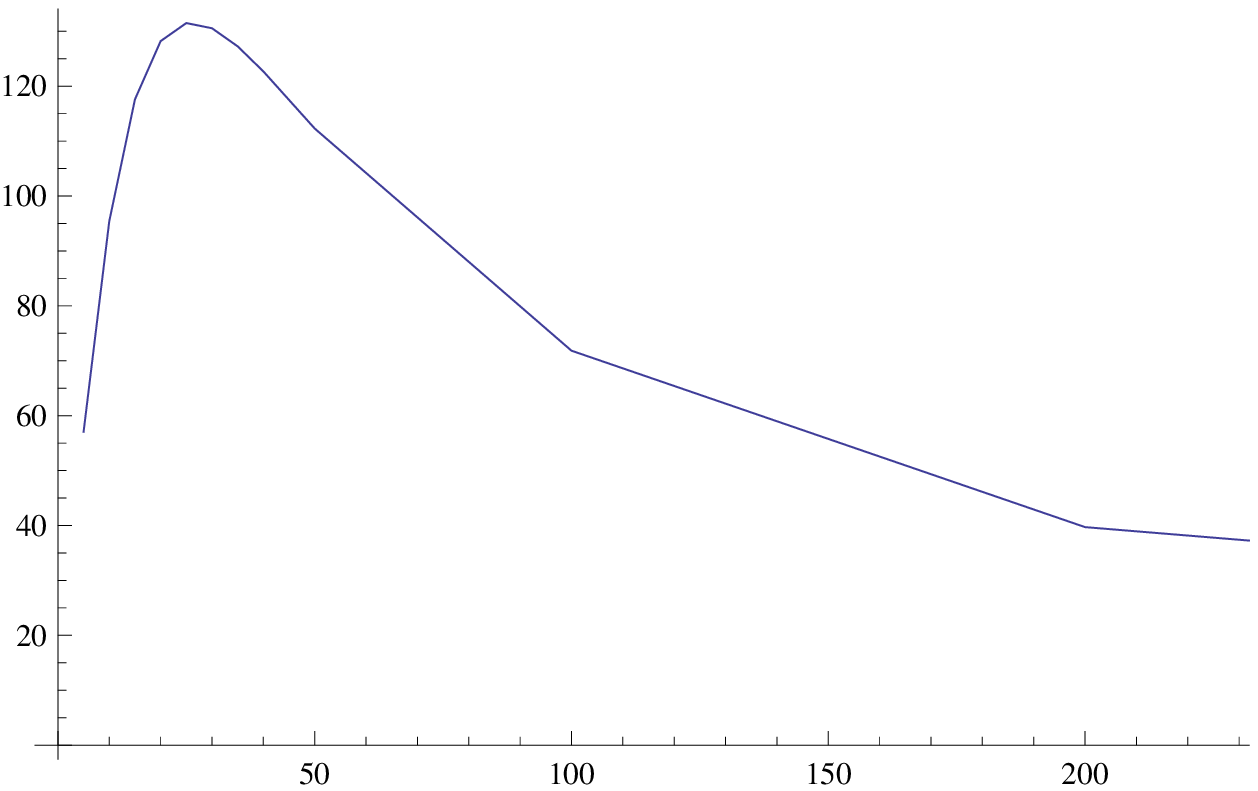}
}
\subfloat[]
{

\rotatebox{90}{\hspace{0.0cm} $\left . R_0\right|_A\rightarrow$events/(kg-y)}
\includegraphics[width=0.45\textwidth]{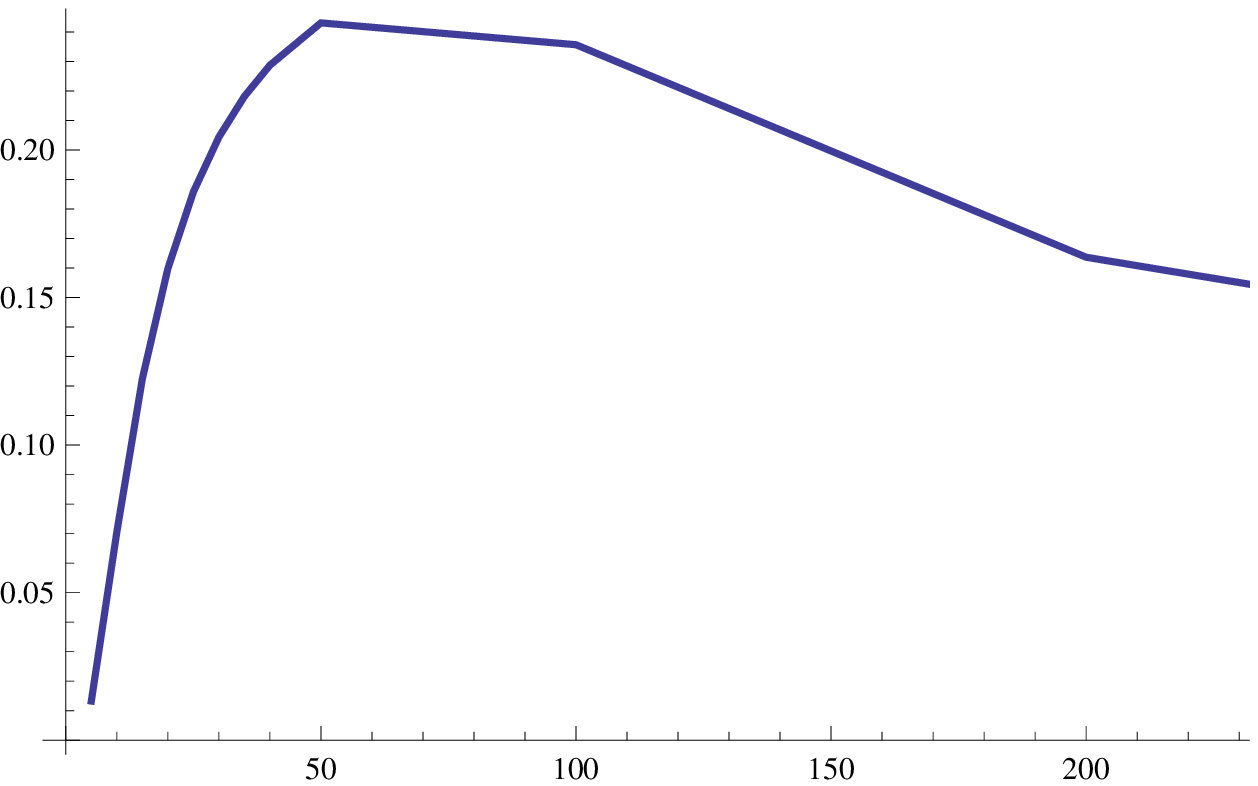}
}
\\
{\hspace{-2.0cm} $m_{\mbox{{\tiny WIMP}}}\rightarrow$GeV}\\
\caption{ We show for the target $^{83}$Kr  the time average total rate   as a function of the WIMP mass in GeV obtained in the case of elastic scattering (a) and inelastic scattering (b), assuming a zero energy threshold. Note that the maximum in the case of the inelastic scattering occurs  at a higher mass compared to the location of the maximum for the ground state transition.
 \label{fig:totR0Kr}}
\end{center}
\end{figure}

The total rates obtained assuming zero energy threshold are exhibited in Fig. \ref{fig:totR0Kr} as functions of the WIMP mass.

 From this figure we see that the maximum total rate for the inelastic transition is around a WIMP mass of about 50 GeV, while the location of the maximum for ground state transition is around 20 GeV. On the other hand the excited state cannot be populated in the case of WIMPs lighter than 4.6 GeV, but the ground state can, albeit with small rate there. We also see that for a WIMP mass of 20 GeV the obtained rate is $0.16$ counts/(kg-y)$=4.4\times 10^{-4}$/(kg-d). This is just above the rate of  $2.8 \times 10^{-4}$ counts/(kg-d) reported previously \cite{EFL88} . This is a little surprising  because a larger spin $\left |ME \right |^2$ with the value of $0.8\times (0.08/1.91^2)=0.0175$ was estimated in the earlier work. The larger rate obtained in the present work could be attributed to our use of an energy dependent structure function, or, most likely, to  
 the supersymmetric model they used, which  would have yielded   a nucleon cross section smaller than the one used here.

\section{ Experimental aspects of inelastic nuclear scattering rates} 
In this section, we discuss experimental aspects of SD WIMP studies by measuring inelastic nuclear scattering. So far SD and SID WIMPs have been studied experimentally by measuring nuclear recoils of elastic scatterings. 

SD WIMPs may show fairly appreciable cross sections of inelastic spin excitations, as shown in previous sections. Experimentally, inelastic nuclear excitations provide unique opportunities
for studying SD WIMPs.  Inelastic excitations are studied in two ways: singles measurement of both the nuclear recoil energy $E_R$ and the decaying $\gamma $-ray energy $E_{\gamma }$ in one detector, and  coincidence measurement of the nuclear recoil and the $\gamma $-ray in two separate detectors. This is done in a fashion analogous to that of the earlier analysis \cite{VerEjSav13}. So far SD and SID WIMPs have been studied experimentally by measuring nuclear recoils of elastic scatterings. 
 Experimentally, inelastic nuclear excitations provide unique opportunities
for studying SD  rates for WIMP-nuclear interactions.  Inelastic excitations are studied by two ways,  A: singles measurement of both the nuclear recoil energy $E_R$ and the decaying $\gamma $-ray energy $E_{\gamma }$ in one detector, and B: coincidence measurement of the nuclear recoil and the $\gamma $-ray in two separate detectors. The merits of each of them are as follows.
 
\begin{figure}
\begin{center}
\subfloat[]
{
\includegraphics[width=0.45\textwidth]{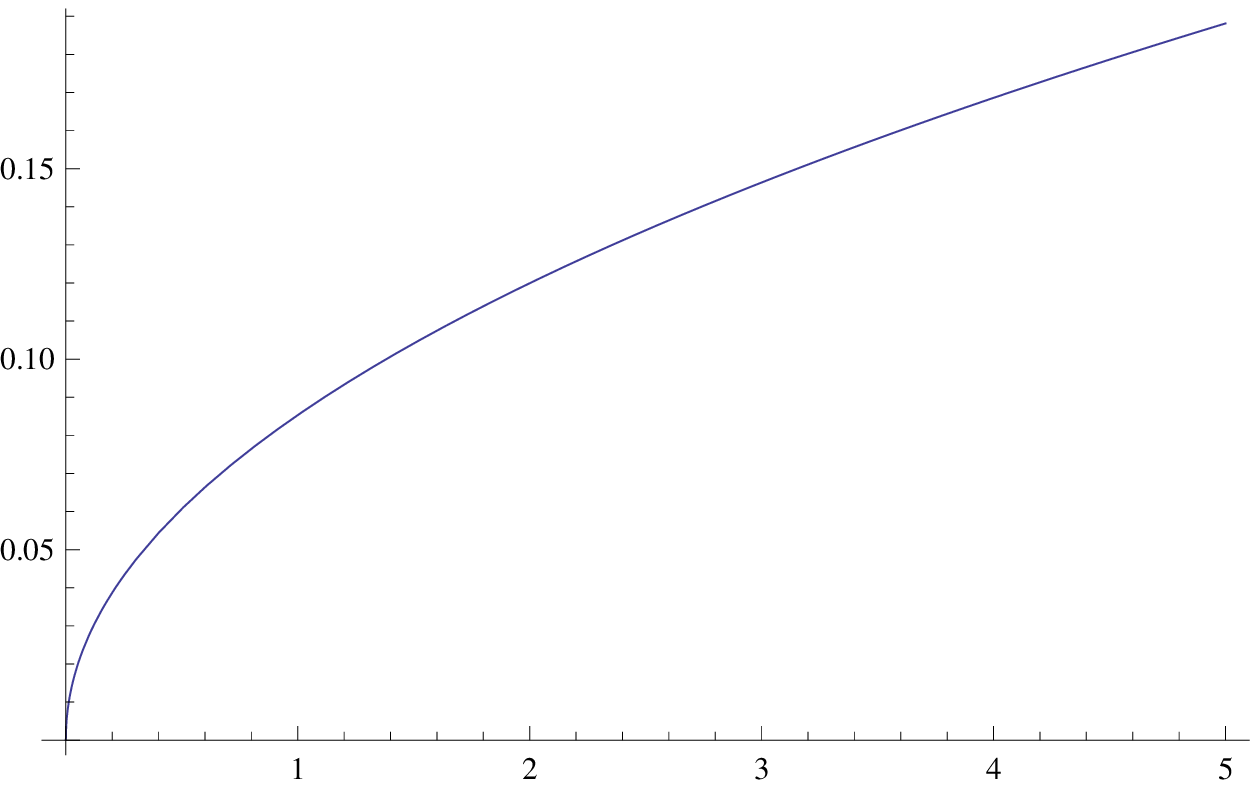}
}
\subfloat[]
{
\includegraphics[width=0.45\textwidth]{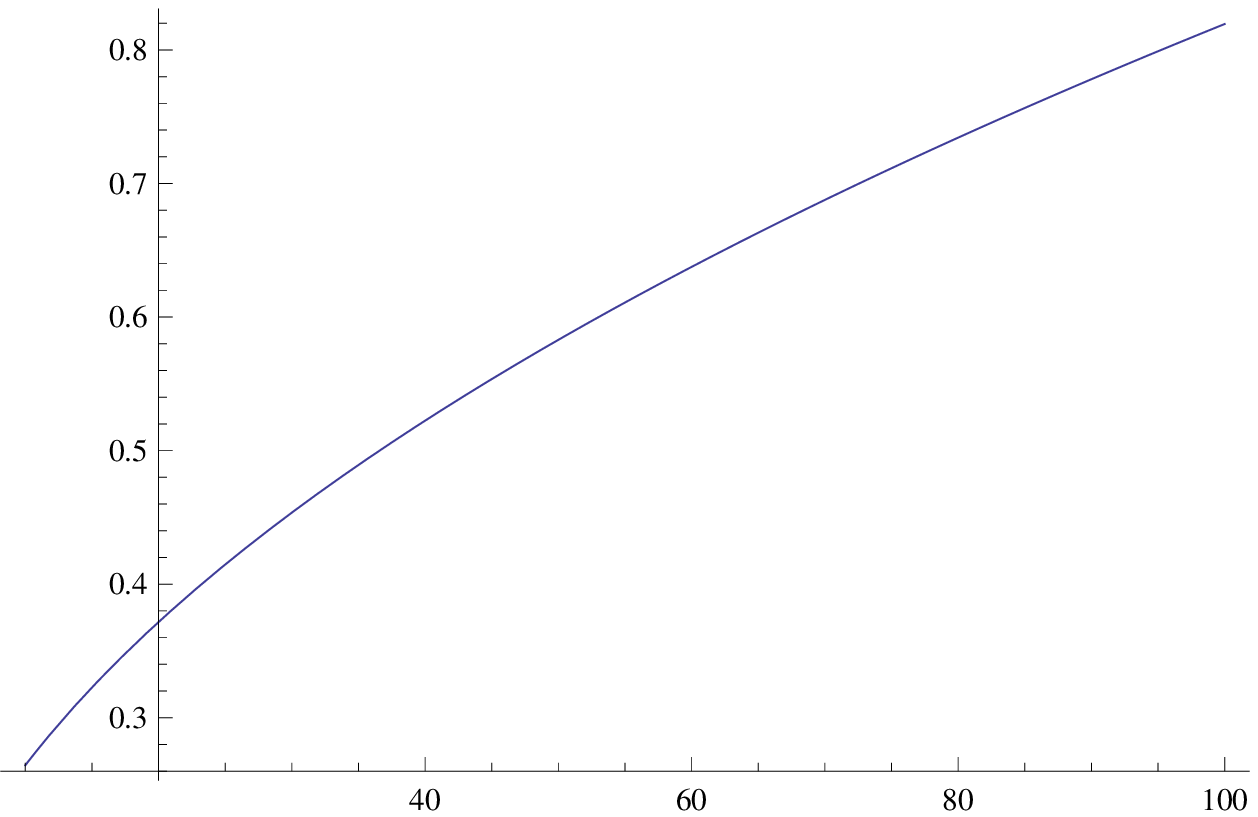}
}
\\
{\hspace{-2.0cm} $E_{R}\rightarrow$keV}
\caption{ The quenching factor as a function of the recoil energy in keV obtained in the Lindhard theory is presented at low energy (a) and at somewhat larger energies (b).}
\label{fig:Quenchf}
\end{center}
\end{figure}

 The large energy signal is obtained by summing the nuclear recoil signal and the $\gamma $ ray signal. It is given as  
\begin{equation}
E({\rm ex})=E_{\gamma} + Q(E_R({\rm ex}))E_R({\rm ex}), 
\end{equation}
where $E_R({\rm ex})$ is the nuclear recoil energy,  $E_{\gamma }$ is the excitation
energy and $Q(E_R({\rm ex}))$ is the quenching factor for the recoil energy signal. In most scintillation and ionization detectors, the quenching factor is as small as $Q(E_R({\rm ex})) \approx 0.1 - 0.05$. It must be determined for each target and detector experimentally. For $^{83}$Kr, the primary target considered in this work, the quenching factor  is not important for a gaseous detector,  but for a liquid or solid  it is taken to be $0.08-0.1$.  An overall picture can be obtained by a phenomenological approach based on the Lindhard theory
 \cite{Lindhard,Simon03}, which is exhibited in Fig. \ref{fig:Quenchf}. In the case of $^{83}$Kr, due to the low excitation energy, our conclusions about the inelastic transition are not affected much by the quenching factor. In the present work we will use a quenching factor of 0 0.08.

Therefore it appears that the energy deposited is mainly the excitation energy. This is much larger than just the recoil energy signal of $E({\rm gr}) =Q(E_R({\rm gr}))$, which is quenched, depending on the detector. 

The sharp rise of the energy spectrum at the 
energy of $E_{\gamma} + Q(E_{min}) E_{min}$, where $E_{min}$ is the minimum energy transfer to the recoil nucleus.  This makes it possible to identify the WIMP nuclear interaction.  On the other hand, the recoil energy spectrum $E_R({\rm gr})$
is continuum like back ground at the low energy region, and thus is hard to be identified.

$E(ex)$ is well above the detector threshold $E(th)$, while
the main part of $E({\rm gr})$ is cutoff by $E(th)$. Accordingly, the event rate $R({\rm ex})$ is about the same order of magnitude as $R(gr)$ for SD WIMPs although the inelastic cross section is much smaller than the elastic one. \\
\begin{figure}
\begin{center}
\subfloat[]
{
\rotatebox{90}{\hspace{0.0cm} $\left .\left (\frac{dR_0}{dE_R}\right )\right|_A \rightarrow$events/(kg-y)/ keVee}
\includegraphics[width=0.45\textwidth]{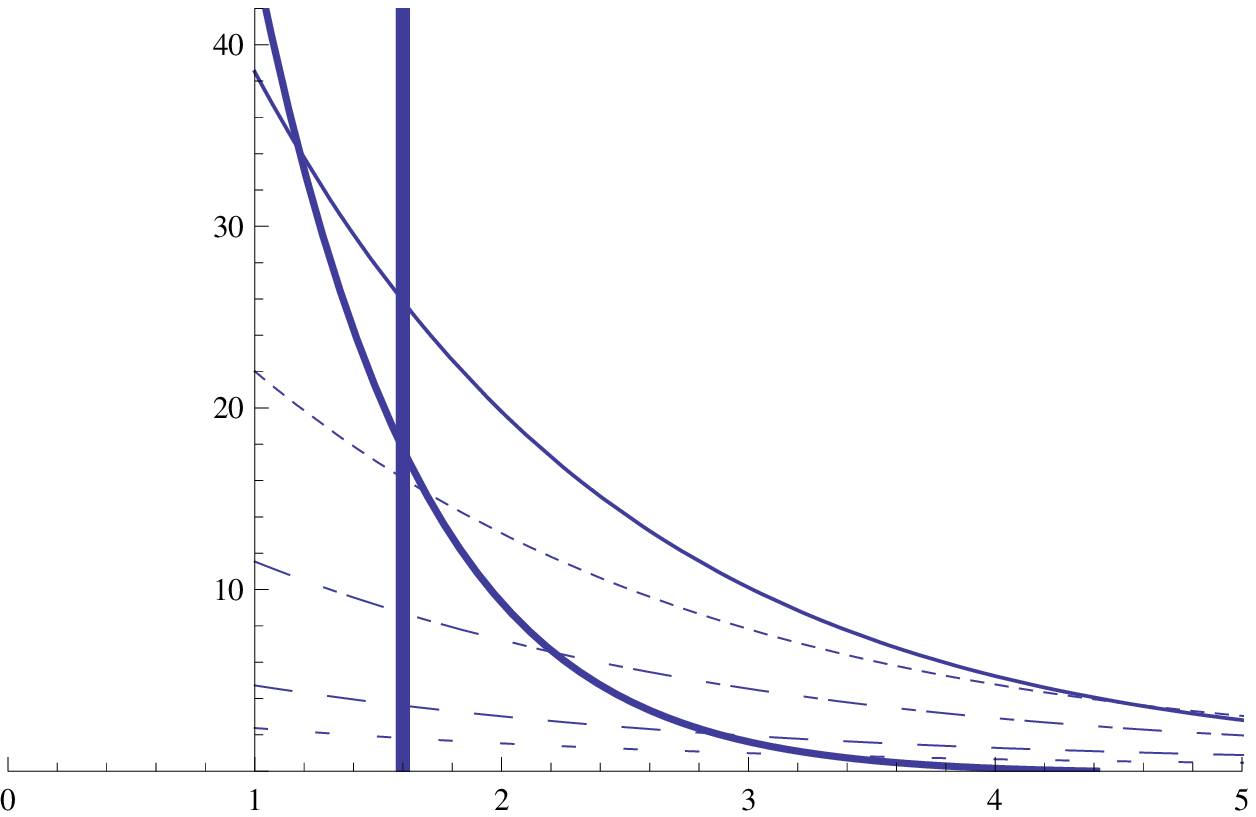}
}
\subfloat[]
{
\rotatebox{90}{\hspace{0.0cm} $\left .\left (\frac{dR_0}{dE_R}\right )\right|_A \rightarrow$events/(kg-y)/ keVee}
\includegraphics[width=0.45\textwidth]{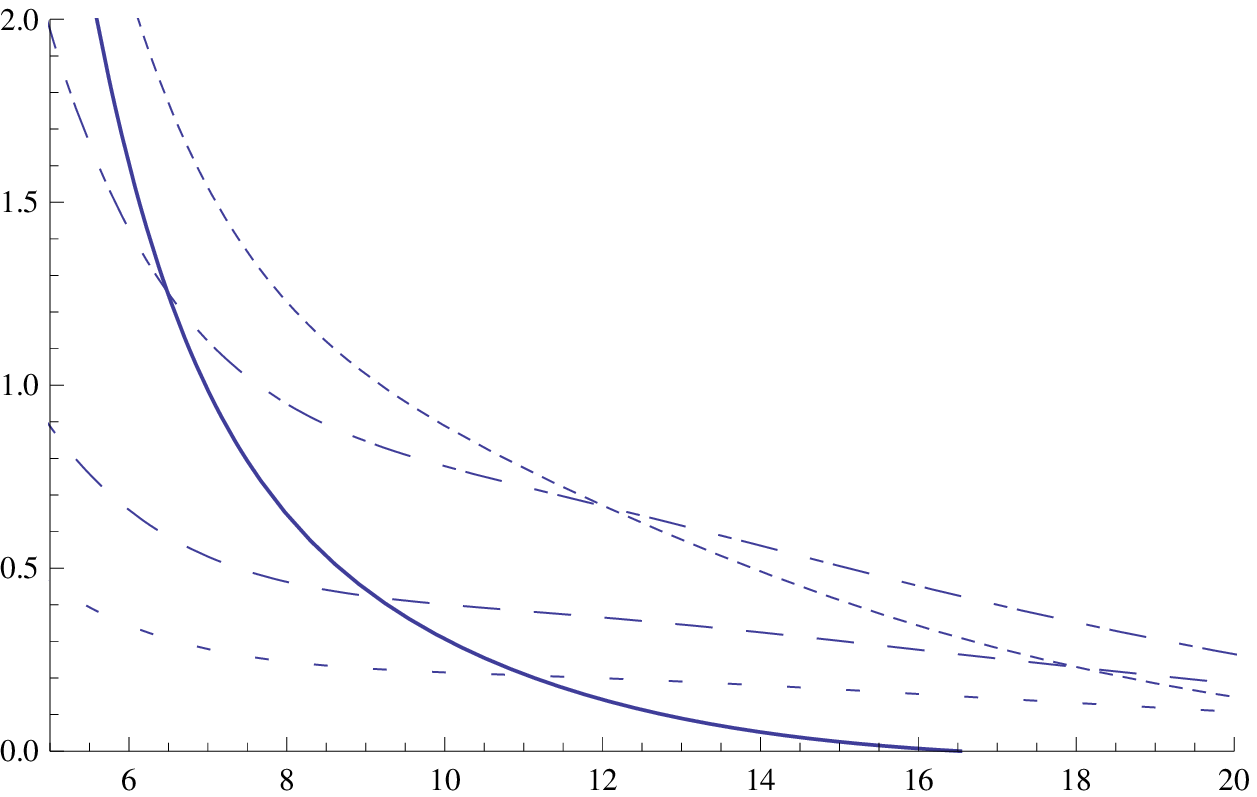}
}\\
\subfloat[]
{
\rotatebox{90}{\hspace{0.0cm} $\left .\left (\frac{dR_0}{dE_R}\right )\right|_A \rightarrow$events/(kg-y)/ keVee}
\includegraphics[width=0.45\textwidth]{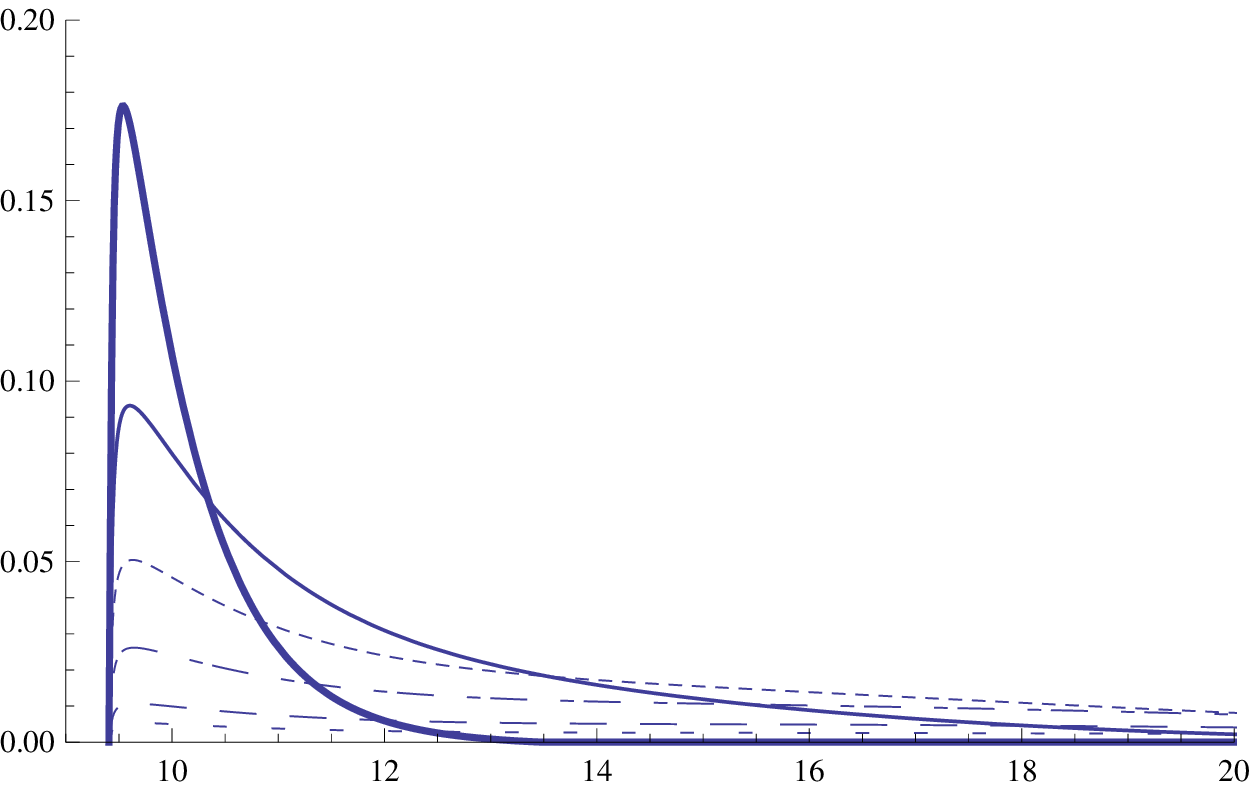}
}
\subfloat[]
{
\rotatebox{90}{\hspace{0.0cm} $\left .\left (\frac{dR_0}{dE_R}\right )\right|_A \rightarrow$events/(kg-y)/ keVee}
\includegraphics[width=0.45\textwidth]{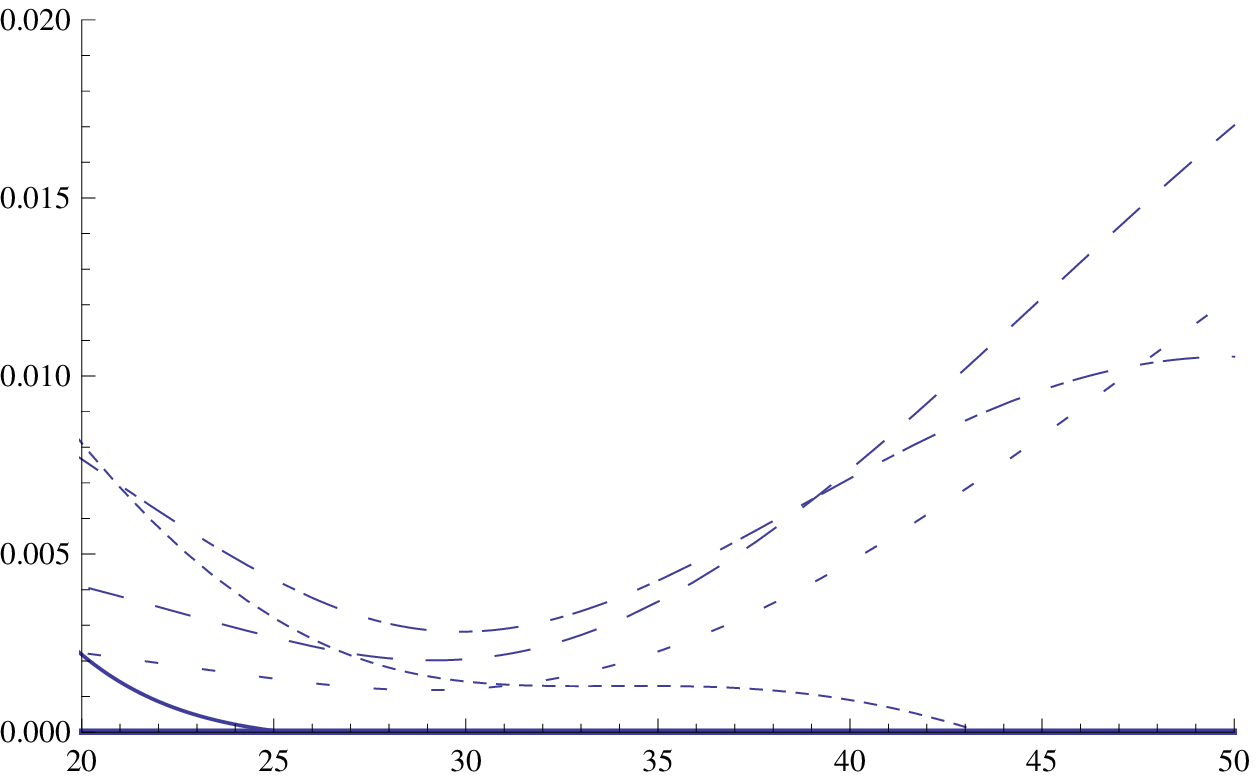}
}
\\
{\hspace{-2.0cm} $E_{R}\rightarrow$keVee}
\caption{ Energy spectrum  for WIMP $^{83}$Kr elastic scattering (panels (a) and (b)) in two energy regions. Note that  the maximum allowed energy transfer depends on the WIMP mass. For a heavy WIMP it can reach  about 60 keVee, but it falls real fast with the energy transfer. The energy spectrum  for the 9.4 keV excited state inelastic scattering is also 
shown (panel (c)). The quenching factor employed was $Q$=0.08 and the energy threshold $E(th)$ = 1.6 keV. The x and y scales are the electron-equivalent energy and the rate per unit electron equivalent energy.}
\label{fig:Kr83}
\end{center}
\end{figure}
The typical energy spectra to be measured experimentally for the elastic and inelastic transitions of $^{83}$Kr  are shown in Fig. \ref{fig:Kr83}.  Here we assumed detectors with the quenching factor of $Q$=0.05 and the energy threshold of $E(th)$ = 1.6 keV. The yield on the y axis is the one per unit energy of the electron equivalent energy, i.e. $QE_R$ and the energy on the x axis is the electron equivalent energy. 

We note that the yield is enlarged by a factor $1/Q$=20, while the energy is reduced by a factor $Q$=0.05. The low energy part of the elastic scatting is cut off by the threshold energy of 1.6 keV electron equivalent energy, i.e. 32 keV recoil energy.

\begin{figure}
\begin{center}
\subfloat[]
{
\includegraphics[width=0.45\textwidth]{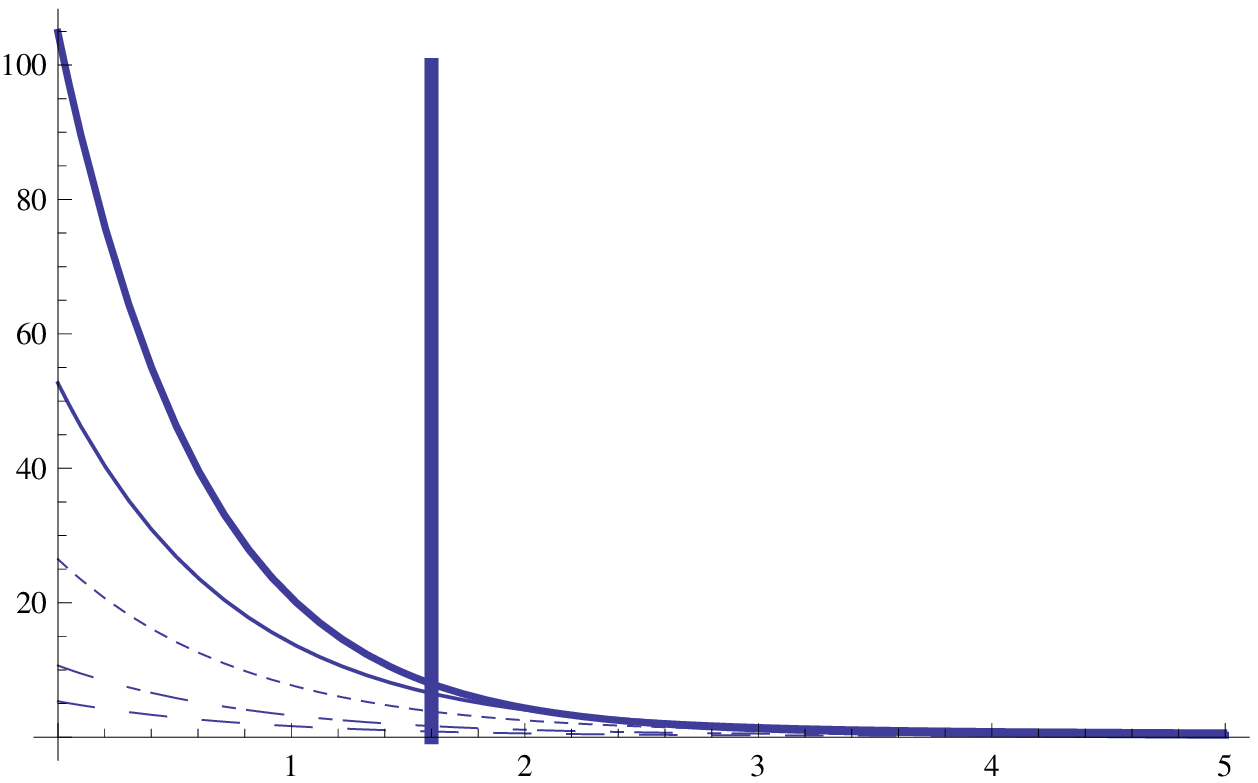}
}
\subfloat[]
{
\includegraphics[width=0.45\textwidth]{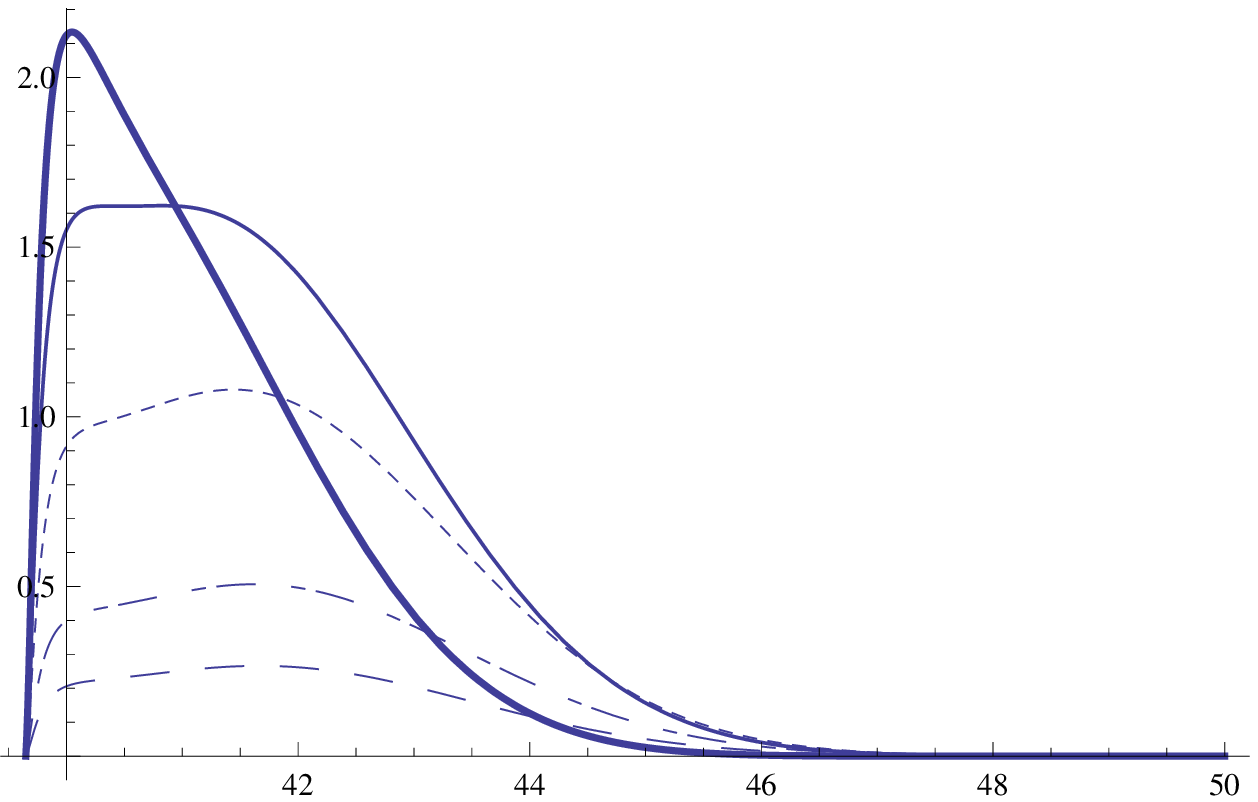}
}
\\
{\hspace{-2.0cm} $E_{R}\rightarrow$keVee}
\caption{ We show the energy spectrum for the elastic transition (a) and the inelastic to the first excited state (b) for the target $^{129}$Xe}.
\label{fig:Xe129}
\end{center}
\end{figure}

\begin{figure}
\begin{center}
\subfloat[]
{
\includegraphics[width=0.45\textwidth]{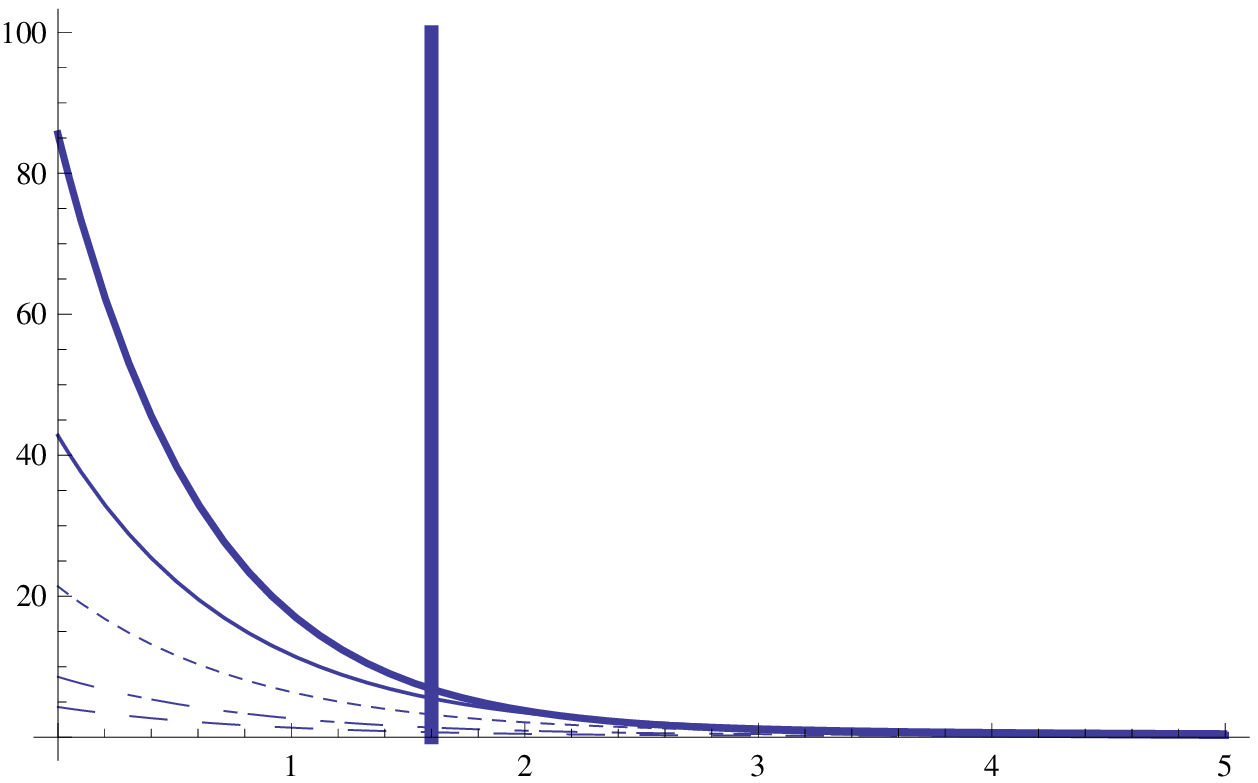}
}
\subfloat[]
{
\includegraphics[width=0.45\textwidth]{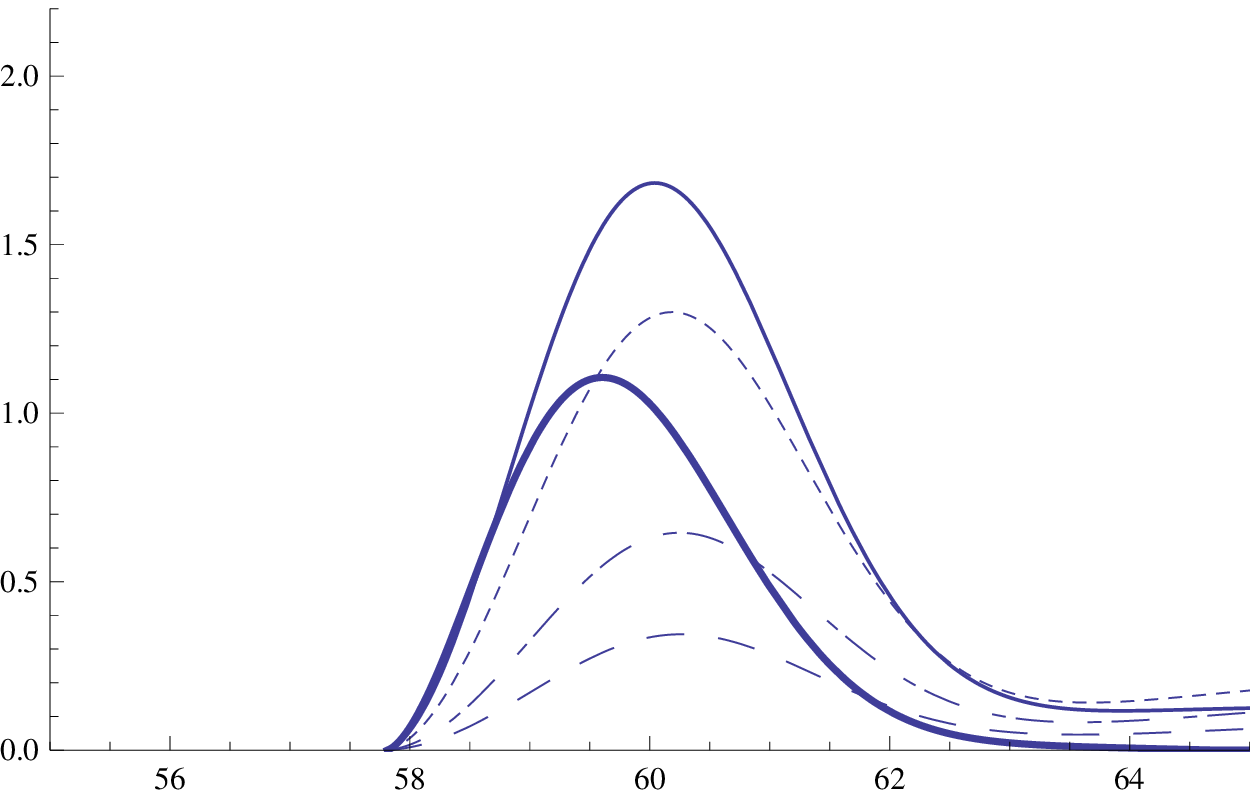}
}
\\
{\hspace{-2.0cm} $E_{R}\rightarrow$keVee}
\caption{ The same as in Fig. \ref{fig:Xe129} for the target $^{127}$I}.
\label{fig:I127}
\end{center}
\end{figure}

For comparison we plot the same data for Xe $^{129}$Xe  and $^{127}$I in Figs \ref{fig:Xe129} and \ref{fig:I127} respectively , which have appeared elsewhere \cite{VerEjSav13}.

Comparing the two targets we see that the smallness of the spin ME in the case of $^{83}$Kr is  compensated by the favorable energy  dependence of its inelastic spin structure function and the low excitation energy.
\section{Experimental Feasibility}
       The main purpose of the research described in this article is to explore the feasibility of an experiment to search for Cold Dark Matter via an inelastic excitation of a nuclear target. The isotope $^{83}$Kr was chosen for two reasons: first, it has a low energy M1 transition from the 7/2$^+$ first excited state at 9.4-keV to the 9/2$^+$ ground state; second, in the liquid state it is a fairly good scintillation detector material. The disadvantage of this choice is that the natural abundance of $^{83}$Kr is only 11.5$\%$ which will require isotopic enrichment. In addition any content of the ubiquitous radioactive isotope $^{85}$Kr will render a sensitive search ineffective. Even the isotopic enrichment of Kr gas from the usual sources will never be free enough from $^{85}$Kr to allow a sensitive experiment. Fortunately, a recent discovery by other researchers can very probably be used to ameliorate this difficulty. First, we discuss the scintillation properties of liquid Kr, and then possible new sources of low-radioactive Kr gas.
      The scintillation and other physical properties of liquid Kr has been well studied and published in the literature \cite{ABC93}- \cite{TDM99}. It has been shown by independent groups \cite{ABC93} \cite{ABBD06} that the addition of a few percent by mass of Xe enhances the fast component of scintillation of LKr by a factor of approximately 10. Considering the long lifetime of the 7/2$^+$ to 9/2$^+$ ground-state decay in $^{83}$Kr, the pulse shape distortion might be utilized to provide partial background rejection as discussed in another case \cite{Avignone00}. Nevertheless, this advantage would be minimal in the case that the background from $^{85}$Kr was not negligible. 
      In the usual case, sources of krypton and argon for industrial use are obtained by distillation of atmospheric air, which contains radioactive $^{85}$Kr (10.75y) and $^{39}$Ar (269y). These contaminations are unacceptable for use in ultra-low background experiments. It is a well-known fact in the geology and geochemistry community that through petroleum exploration, very large deep underground accumulations of CO$_2$ have been discovered in the greater Colorado Plateau and in the Southern Rocky Mountain Region \cite{Burnard12}. Recently the Princeton-lead Darkside Collaboration has been successful in working with industry to extract ultra-low radioactive background argon from large volumes of CO$_2$ using a Vacuum Pressure Swing Absorption plant \cite{Xu12},\cite{Back15}. In the case of Kr, this would be the input gas for the isotopic enrichment. 
      These facts imply the possibility of a sensitive search for the Cold Dark Matter excitation of $^{83}$Kr to the first excited state and the detection of the deexcitation 9.4-keV gamma ray. However, the extraction of the Kr from the deep underground CO$_2$, as well as the requirement of isotopic enrichment would be far more costly than the extraction of $^{40}$Ar, which has an isotopic abundance of 100$\%$. Accordingly, a research and development study would have to be undertaken to determine the technical and cost feasibility of this experimental approach. However, of the 30 isotopes discussed by Ellis, Flores and Lewin \cite{EFL88} $^{83}$Kr is one of the few for which there are clear paths to building a large detector.           	

\section{Concluding remarks}

 SI and SD WIMPs have extensively been studied, so far, by measuring elastic nuclear recoils.  The elastic scattering of SI WIMPs is coherent scatting, thus the cross section is enhanced by the factor $A^2$ with $A$ being the nuclear mass number.  On the other hand the elastic cross section of SD WIMPs is, is in general, smaller by 2-3 orders of magnitude than that for SI WIMPs because the spin induced rates do not depend on $A^2$, i.e. they do not exhibit coherence. It may, however, compete with the coherent scattering in models in which the spin induced nucleon cross section is much larger than the one due to a scalar interaction. We have seen that there exist viable such particle models. In such cases the inelastic WIMP-nucleus scattering  becomes important.
 
 Indeed the inelastic scattering via spin interaction provides a new opportunity for studying SD WIMPs. Experimentally, observation of both the nuclear recoils and the $\gamma $ ray following the excited state does lead to the large energy signal of the unquenched $E_{\gamma }$ and  the sharp rise of the energy spectrum at around $E_{\gamma }$.  Even though the SD inelastic cross section is smaller than the SD elastic one, the inelastic event rate is comparable with the elastic one, since the inelastic signal is well  beyond the detector threshold energy, while the elastic signal is mostly cut off by the detector threshold. 

In the present paper we discussed mainly the inelastic excitations of $^{83}$Kr. Another possible isotope, in addition to the $^{127}$I and $^{129}$Xe discussed previously \cite{VerEjSav13}, is  the $^{73}$Ge, in high energy resolution Ge detectors, and $^{125}$Te currently under study. 
In short, the present paper, in conjunction with the earlier calculations \cite{VerEjSav13},  indicates that the inelastic scatting opens a new powerful way to search for SD WIMPs.

{\bf  Acknowledgments}: One of the authors (JDV) is indebted to H. Ejiri for enlightening discussions and his comments, especially on the quenching factors, and the Physics Department of the University of S. Carolina for their support and hospitality.
The authors are grateful to Frank Calaprice for very helpful discussions. This work was partially  supported by the Academy of Finland under the Centre of Excellence Programme 2012-2017 (Nuclear and Accelerator Based Physics Programme at JYFL) and the FIDIPRO program. JDV was supported by a USC Provost’s Internal Visiting Distinguished Visitor Grant and
FTA was supported by National Science Foundation Grant NSF PHY-1307204.


\end{document}